\documentclass[10pt,a4paper]{article}

\usepackage[utf8]{inputenc}
\usepackage[T1]{fontenc}
\usepackage{authblk} 
\usepackage{geometry}
\usepackage{amsmath, amssymb, amsfonts}
\usepackage{subcaption}
\usepackage{graphicx}
\usepackage{float}
\usepackage{titlesec}
\usepackage{cite}
\usepackage[colorlinks=true, allcolors=blue]{hyperref} 
\titlespacing*{\paragraph}{0pt}{6pt}{6pt}

\newcommand{\e}{\mathrm{e}}
\newcommand{\ii}{\mathrm{i}}

\title{\textbf{Cavity Solitons as a Nonlinear Substrate for Photonic Neuromorphic Computing}}

\author[1,2,*]{Amir Arsalan Arabieh}
\author[1,2]{Alessandro Lupo}
\author[2]{Simon-Pierre Gorza}
\author[1]{Serge Massar}

\affil[1]{\small Laboratoire d’Information Quantique, CP 224, Université libre de Bruxelles (ULB), Av. F. D. Roosevelt 50, 1050 Brussels, Belgium}
\affil[2]{\small Service OPERA-Photonique, CP 194/5, Université libre de Bruxelles (ULB), Av. F. D. Roosevelt 50, 1050 Brussels, Belgium}
\affil[*]{\small Corresponding author: amirarsalan.arabieh@ulb.be}

\date{} 

\begin{document}
	\maketitle

\begin{abstract} 
	Reservoir computing leverages nonlinear dynamics of physical systems to process temporal information with minimal training cost. Here, we demonstrate that cavity solitons sustained in a fiber optical cavity provide an optical platform for photonic reservoir computing. Our methodology exploits the use of a phase-modulated drive laser to encode the input, while the reservoir states are accessed through frequency-resolved readout. Numerical simulations indicate that the emission of Kelly waves enriches the dynamics and enhances performance for machine learning tasks. We evaluate the performance of the cavity-soliton reservoir computer on several standard benchmark tasks.
\end{abstract}

\section{Introduction}
Hardware implementations of bio-inspired machine learning algorithms are receiving increasing attention because of their potential for low energy consumption and the fundamental new insights they provide into the dynamics of nonlinear dynamical systems~\cite{Schuman2022, Shastri2021}. Among the algorithms studied, Reservoir Computing (RC)~\cite{jaeger2004,Maass2002,Verstraeten2007}, a machine learning paradigm for processing time-dependent information, has attracted much attention due to its good performance on many benchmark tasks, ease of training, and the large variety of physical systems in which it can be implemented. As shown in the schematic of Fig.~\ref{fig:RCpanel}, a reservoir computer consists of an input layer, a reservoir, and an output layer. The reservoir is a recurrent nonlinear dynamical system that maps the time-dependent inputs into a higher-dimensional space, where patterns that are not initially linearly separable become so, allowing for efficient training. The short-term memory property~\cite{Yeager_shortterm} ensures that the influence of past inputs on the reservoir's dynamical state decays over time, and hence that only recent inputs influence the output layer. Due to its simple architecture, many physical systems have been proposed as reservoirs (see Refs~\cite{Tanaka2019,Yan2024} for reviews). In particular, optical systems have been studied extensively, motivated by the potential for energy efficiency, parallelism, and scalability on integrated chips. Photonic approaches~\cite{VanderSande2017} include optoelectronic delay-based systems~\cite{Appeltant2011}, lasers with optical feedback~\cite{Shen2023}, and frequency-based systems in which the frequency degree of freedom of the optical field is exploited~\cite{Butschek2022} (see Ref~\cite{Abreu2024} for a review).

Independently of the above work, optical cavity solitons (CSs)~\cite{Akhmediev2005, Akhmediev2008, Hasegawa1989} have been the subject of extensive study both because of their fundamental scientific interest and their potential for applications, in particular as sources of optical frequency combs~\cite{Suh2016, Yu2018, Bao2021, Palomo2017, Geng2022}. Moreover, they have been demonstrated in fiber-optics cavities~\cite{Leo2013, Englebert2021}, and in microring resonators~\cite{Herr2014, Kippenberg2018}. CSs are optical pulses phase-locked to the driving field and arise from a double-balance: the pulse broadening due to chromatic dispersion is compensated by self-phase modulation, and intrinsic cavity losses are compensated by continuous energy injection from a coherent external driving field. As a result, the characteristics of CSs, such as their temporal width and peak power, are determined by the system parameters, i.e., the optical-cavity properties and cavity detuning.   

The potential of cavity solitons for information processing has been studied previously. Their precisely controllable temporal positions make them suitable for use as optical memories~\cite{McDonald1990,Pimenov2020}. Integrated soliton microcombs have been used for matrix-vector multiplication~\cite{Feldmann2021, Xu2021, Xu2020}. They have been used for solving combinatorial optimization problems via an Ising model representation~\cite{Jin2025}. The concept of leveraging nonlinear wave propagation for neuromorphic computing was first put forward in Refs.~\cite{Marcucci2020, Zhou2022}, and a first proposal for using conservative Kerr solitons for reservoir computing, albeit challenging to implement in practice, was suggested in~\cite{Silva2021}.

Recently, frequency multiplexing has emerged as a very promising method to leverage linear and nonlinear wave propagation for neuromorphic computing both in the context of extreme learning machines~\cite{Lupo2021, Hary2025, Saeed2025, Zajnulina2025} and reservoir computing~\cite{Li2023, Skontranis2023, Cox2025}. Cavity solitons represent a particularly compelling physical substrate for this approach. First, their naturally broad and coherent optical spectra provide a high-dimensional state space ideal for wavelength-division multiplexing. Second, the inherent nonlinear response of a CS to external perturbations provides the necessary activation function for complex information mapping. Finally, as stable dissipative structures and attractors of the system, CSs ensure reproducibility of the reservoir states and short-term memory property.

Reservoir computing based on frequency-comb dynamics has been explored through numerical studies of chaotic Kerr cavities~\cite{Shishavan2025}, as well as the pioneering experimental demonstration of reservoir computing using cavity-soliton frequency combs which was recently reported in microring resonators~\cite{Cuevas2025}. Our work explores this concept in a complementary physical regime using a fiber-based resonator. We demonstrate both numerically and experimentally that the transient dynamics of cavity solitons can serve as a nonlinear substrate for photonic neuromorphic computing, and in particular for reservoir computing. To this end, we encode the time-dependent input signal in the phase of the driving laser, as illustrated in Fig.~\ref{fig:CSRC}. This perturbation disrupts the soliton, which tends to relax back to its stationary state via  transient dynamics. In the frequency domain, the soliton spectrum undergoes a complex, time-dependent breathing behavior. As illustrated in Fig.~\ref{fig:specbreath}, the reservoir nodes are defined by dividing the optical spectrum into distinct frequency channels, with the time-dependent spectral power in each channel used as a reservoir state. 

Furthermore, the present study explores the role of Kelly sidebands~\cite{kelly1,kelly2} in the performance of soliton-based reservoir computing. When solitons encounter a localized perturbation, they emit radiative waves. Owing to cavity periodicity, radiation emitted over successive roundtrips adds coherently and appears in the frequency domain as sidebands on the soliton spectrum, known as Kelly sidebands (experimental and numerical demonstrations of this phenomenon are shown in Fig.~\ref{fig:expsetup_spec} and Fig.~\ref{fig:Ikeda1_evol}, respectively); while in the time domain they form oscillatory tails attached to the soliton~\cite{Wang2017}. To study the contribution of Kelly sidebands, we numerically benchmark the reservoir computing performance by comparing the predictions of three theoretical frameworks describing the nonlinear dynamics of cavity solitons as a reservoir. The most realistic model is the lumped element model (also known as the Ikeda map~\cite{ikeda1979multiple}) which describes the intracavity evolution of the field at each roundtrip followed by the boundary condition at the coupler. This can be simplified to the mean-field Lugiato--Lefever Equation (LLE)~\cite{lugiato1987spatial,coen2012modeling,Haelterman1992,Chembo2013}. Despite its simplicity, it is a universal model that predicts remarkably well the dynamics of CSs both in microrings and in fiber cavities regardless of the large size difference. The LLE model can be further simplified to a reduced model~\cite{Anderson1983,Hasegawa2002,Herr2016-2} consisting of two coupled differential equations for the soliton amplitude and phase. Unlike the mean-field LLE and the reduced models, which are invariant under the fast-time translation, the discrete roundtrip nature and boundary conditions of the Ikeda map break this temporal symmetry, thereby providing the periodicity required for the emergence of Kelly sidebands. It is worth emphasizing that a previously related study~\cite{Cuevas2025} did not investigate this aspect of soliton dynamics. In the weak-coupling, near-uniform-parameter limit, the mean-field LLE provides an appropriate description. By contrast, our active fiber cavity, similar to~\cite{Englebert2023} but operated at a considerably larger coupling ratio—readily excites Kelly sidebands in the soliton spectrum, making the Ikeda map the more suitable modeling framework.

We experimentally demonstrate the resulting reservoir computer on benchmark tasks including the XOR classification task and the Hénon map prediction. We also study the cavity-soliton-based reservoir computer using numerical simulations. The simulations are performed in a more idealized, noise-free environment, which allows us to investigate in more detail the potential of this system. We show that it can in principle reach performance comparable to other experimental systems on several complex benchmarks. 

The paper is organized as follows. In Section~\ref{sec:RC}, we review the principles of reservoir computing, followed by Section~\ref{sec:CS_RC}, which details the utilization of cavity solitons as physical reservoirs. Section~\ref{sec:models} provides an overview of the theoretical models describing the dynamical evolution of cavity solitons. Our detailed numerical simulations are presented in Section~\ref{sec:NS}: Section~\ref{sec:modelCompare} compares the predictions of different models when leveraged for the reservoir dynamics, while Section~\ref{sec:RCbench} benchmarks the system against standard reservoir computing tasks. Our fiber-based experimental implementation is presented in Section~\ref{sec:EXP_setup}, with the corresponding results discussed in Section~\ref{sec:EXP_results}. Finally, Section~\ref{sec:outlook} provides a discussion and outlook on future directions of this research. The Supplementary Material provides further details on the derivation of the dynamical equations governing the soliton’s amplitude and phase via a variational approach.

\section{Reservoir Computing Framework}
\label{sec:RC}
Reservoir computing (RC) is a recurrent neural network (RNN) paradigm for processing time-series. As illustrated in Fig.~\ref{fig:RCpanel}, at discrete time step $m$, the RC system consists of three layers: (i) an input layer that injects the input $u(m)$ into the network through input weights $\mathbf{W}_{\mathrm{in}}$, (ii) a reservoir with state vector $\mathbf{x}(m)$ of dimension $N$, and (iii) a readout layer. In its standard form, often called Echo State Network \cite{jaeger2004}, the reservoir dynamics is given by  
\begin{align}
	\mathbf{x}(m) &= \mathbf{f}_{\mathrm{res}}
	\Bigl(\mathbf{W}_{\mathrm{res}}\mathbf{x}(m-1) + \mathbf{W}_{\mathrm{in}}\,u(m)\Bigr),
	\label{RC_eq}\\
	\mathbf{x}_{\mathrm{out}}(m) &= f_{\mathrm{out}}\bigl(\mathbf{x}(m)\bigr),
	\label{RC_outnode}\\
	\mathbf{y}(m) &= \mathbf{W}_{\mathrm{out}}\,\mathbf{x}_{\mathrm{out}}(m),
	\label{RC_out}
\end{align}
where $\mathbf{f}_{\mathrm{res} }$ is a nonlinear function, $\mathbf{W}_{\mathrm{res}}$ is the randomly chosen interconnection weights, ${f}_{\mathrm{out} }$ is the output nonlinearity, and $\mathbf{y}(m)$ is the reservoir output. To solve tasks, only the readout weights $\mathbf{W}_{\mathrm{out}} $ are trained, while the internal connections remain fixed.

Consider a sequence of $K$ inputs \(\{u(m)\}_{m=1}^K\), which we divide into a training set and a testing set. The reservoir states and output sequences are stored in matrices whose time dimension has length $K$. Training is performed in a supervised manner: the reservoir state matrix $\mathbf{X}$ is split into a training set $\mathbf{X}_{\mathrm{train}}$ and a test set $\mathbf{X}_{\mathrm{test}}$, and the mean square error (MSE) between the target outputs ${\mathbf{\hat Y}}_{\mathrm{train}}$ and the corresponding reservoir outputs  $\mathbf{Y}_{\mathrm{train}}$ is minimized. The optimal $\mathbf{W}_{\mathrm{out}}$ is obtained via ridge regression:
\begin{equation}
	\mathbf{W}_{\mathrm{out}} =
	\hat{\mathbf{Y}}_{\mathrm{train}}\,\mathbf{X}_{\mathrm{train}}^{\mathsf{T}}
	\left(\mathbf{X}_{\mathrm{train}}\mathbf{X}_{\mathrm{train}}^{\mathsf{T}} + \lambda \mathbf{I}\right)^{-1},
	\label{eq:training}
\end{equation}
where $\lambda$ is the regularization (ridge) parameter. Eq.~\eqref{eq:training} reduces to standard linear regression when $\lambda$ is set to zero. The reservoir performance is often evaluated using the normalized root mean square error (NRMSE)
\begin{equation}
	\mathrm{NRMSE} =
	\frac{\lVert \mathbf{Y}_{\mathrm{test}} - \hat{\mathbf{Y}}_{\mathrm{test}} \rVert_2}
	{\lVert \hat{\mathbf{Y}}_{\mathrm{test}} - \overline{\hat{\mathbf{Y}}}_{\mathrm{test}} \rVert_2},
	\label{eq:NRMSE}
\end{equation}
where $\lVert\cdot\rVert_2$ represents the Euclidean norm and $\overline{\hat{\mathbf{Y}}}_{\mathrm{test}}$ is a vector of the same size as $\hat{\mathbf{Y}}_{\mathrm{test}}$ containing its mean value.

\section{Reservoir Computing with Cavity Solitons}
\label{sec:CS_RC}

\begin{figure}[H]
	\centering
	\begin{subfigure}[t]{0.4\textwidth}
		\centering
		\vspace{0pt}
		\includegraphics[width=\linewidth]{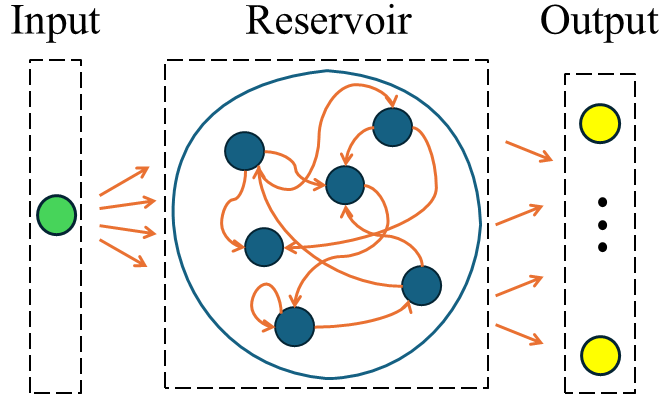}
		\caption{Reservoir Computing Framework}
		\label{fig:RCpanel}
	\end{subfigure}
	\hspace{.5cm}
	\begin{subfigure}[t]{0.4\textwidth}
		\centering
		\vspace{0pt}
		\includegraphics[width=\linewidth]{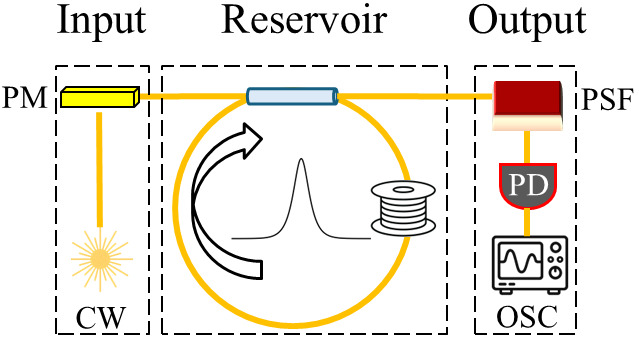}
		\caption{Reservoir computing with cavity solitons}
		\label{fig:CSRC}
	\end{subfigure}
	\begin{subfigure}[t]{0.42\textwidth}
		\centering
		\vspace{0pt}
		\includegraphics[width=\linewidth]{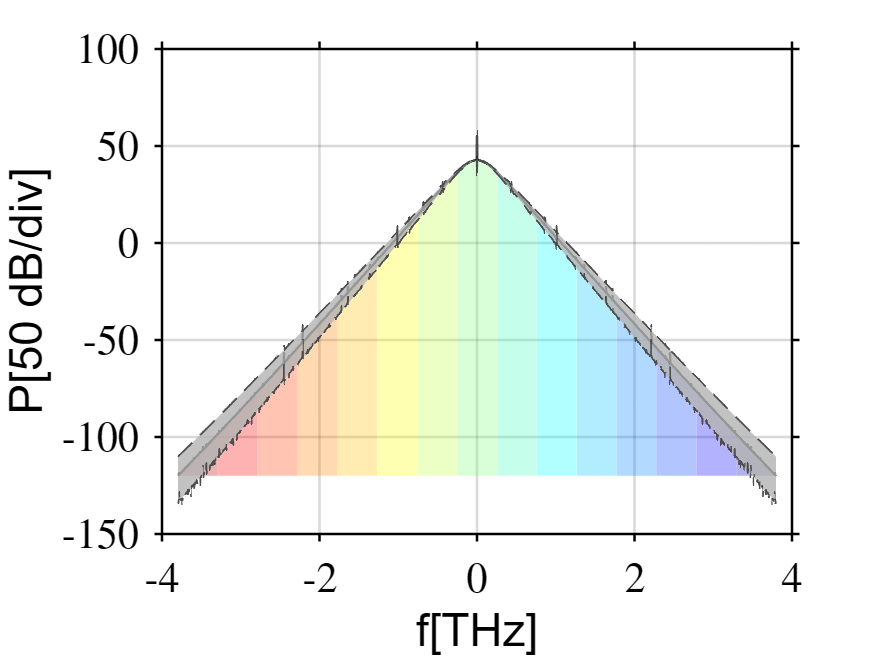}
		\caption{Frequency channels and spectral breathing amplitude in response to phase perturbation}
		\label{fig:specbreath}
	\end{subfigure}
	\hspace{.5cm}
	\begin{subfigure}[t]{0.4\textwidth}
		\centering
		\vspace{0pt}
		\includegraphics[width=\linewidth]{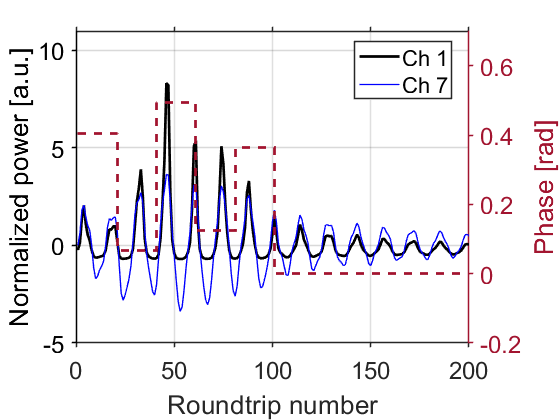}
		\caption{Dynamics of two spectral channels under phase perturbations of the drive}
		\label{fig:relax}
	\end{subfigure}
	\caption{Reservoir computing with cavity solitons.
		(a) Reservoir computing framework showing the input layer, the reservoir, and the readout layer. The weights are indicated in orange.
		(b) Implementation of reservoir computing using cavity solitons. The input layer is realized by modulating the phase of a continuous-wave (CW) laser using a phase modulator (PM). The reservoir layer consists in the dynamical evolution of cavity solitons inside the fiber cavity as they respond to the input. The output layer consists of time-dependent measurements of the power in individual frequency channels, using a programmable spectral filter (PSF), a photodiode (PD), and an oscilloscope (OSC). 
		(c) Spectral breathing amplitude of different frequency channels. The colored bands are the frequency bands, positioned at different spectral offsets from the pump frequency, each representing a different reservoir node. The shaded gray area indicates the amplitude of the spectral breathing of the cavity soliton under phase modulation of the driving field. Parameters: $\sigma_{\varphi}\,=\,0.14$, $K\,=\,500$, $q\,=\,10$.
		(d) Temporal dynamics of two distinctive spectral channels in response to five arbitrary input symbols ($q\,=\,20$), as a function of the roundtrip number, after mean subtraction and normalization to their root-mean-square (RMS) values. The black trace corresponds to the most red-detuned channel, while the blue trace corresponds to the channel adjacent to the continuous-wave (CW) pump. The dashed curve indicates the phase modulation applied to the driving laser.
		The spectral breathing and temporal dynamics of two channels are obtained using numerical simulations following the Ikeda map model with constant parameters: $P_{\text{in}}\,=\,150\,\mathrm{mW}$, $\delta\,=\,2\,\mathrm{rad}$, $\Lambda\,=\,0.03$.}
	\label{fig:mainpanel}
\end{figure}

We adapt the general reservoir computing framework described above to cavity soliton dynamics, as shown in Fig.~\ref{fig:CSRC}. The input information is encoded in the phase of the driving laser. To define the $N$ output nodes, frequency multiplexing is employed, with different nodes represented by distinct spectral channels, as illustrated in Fig.~\ref{fig:specbreath}.

The dynamics of the system is parametrized by the driving laser power, the roundtrip losses $\Lambda$, dispersion, Kerr nonlinearity, cavity length, and the cavity detuning $\delta$. Cavity solitons exist only within a specific region of this parameter space, within which the cavity can support zero, one, or multiple solitons. When the phase of the driving field is modulated, the solitons undergo temporal oscillations, which correspond to spectral breathing in the frequency domain (Fig.~\ref{fig:specbreath}). During this process, information mixing arises through Kerr-mediated nonlinear coupling between different frequency channels. 

Because the cavity soliton propagates in a high-finesse cavity, its state evolves slowly from one roundtrip to the next. Therefore, the time scale of the input encoding must be rescaled to match the intrinsic soliton dynamics. To this end, we introduce the \textit{repetition factor} $q$, which defines the number of consecutive roundtrips during which each input symbol $u(m)$ is held constant. The phase of the pump laser at the $n$-th roundtrip is \(\varphi_n\,=\,M\,u(m)\) for \((m-1)q+1 \le n \le mq\), where \(M\) is called the \textit{modulation strength} and governs how strongly the solitons are perturbed by the input. In general, higher values of $M$ are preferred for nonlinear information processing, as they induce a stronger nonlinear response of the soliton, and (in experiment) improve the signal-to-noise ratio (see Section~\ref{sec:EXP_setup}). 
However, cavity solitons are stable attractors of the cavity dynamics, in the sense that they will withstand small perturbations, but will disappear if perturbed too strongly. For this reason, there is a limiting modulation strength $M$ (which depends on $q$ and $\delta$) that the cavity soliton can withstand (see Section~\ref{sec:NS}). Hence, the repetition factor \(q\), the modulation strength $M$, the cavity detuning $\delta$, and the input power $P_{\mathrm{in}}$ are hyperparameters that are adjusted for best performance. Equivalently, we use the standard deviation of the input phase $\sigma_\varphi$, as a task-independent measure of the strength of the driving phase perturbation (see Fig.~\ref{fig:Losses}). 

Figure~\ref{fig:relax} shows the response of two representative frequency channels to a sequence of five input symbols, each held constant for 20 cavity roundtrips ($q\,=\,20$), demonstrating the system’s echo state property as the reservoir dynamics relax toward a steady state while retaining a memory of previous inputs. The output nodes $\mathbf{x}_{\mathrm{out}}(m) $ (see Eq.~\eqref{RC_outnode}) are obtained by measuring the power in different frequency bins, averaged over $q$ successive roundtrips. Reservoir nodes exhibit varying power levels across the frequency spectrum. To preserve the dynamics of each node and prevent high-power nodes from dominating the training, we normalize each output node by its root mean square (i.e., carrying out the normalisation \(\mathbf{X}_{i,:} \;\rightarrow\;
\frac{\mathbf{X}_{i,:}}{\sqrt{\frac{1}{K}\,\lVert \mathbf{X}_{i,:}\rVert_2^2}}\)). The reservoir output is then given by Eq.~\eqref{RC_out} with the output weights computed according to Eq.~\eqref{eq:training}.

\section{Theoretical Models}
\label{sec:models}
Cavity solitons are localized pulses sitting on a CW background and traveling in the resonator~\cite{Lugiato2003}. In this section we present three theoretical frameworks for modelling the nonlinear dynamics of cavity solitons within optical resonators: the Ikeda map~\cite{ikeda1979multiple}, the Lugiato--Lefever equation (LLE)~\cite{lugiato1987spatial,coen2012modeling,Chembo2013}, and a reduced model based on the Lagrangian approach~\cite{Anderson1983,Hasegawa2002,Herr2016-2}.
\subsection{Ikeda Map} 
The Ikeda map, 
\begin{subequations}
	\label{eq:ikeda_map}
	\begin{align}
		\label{eq:nlse}
		\partial_z E^{(n)}(z,\tau) &=
		\left(-\frac{\alpha_{\mathrm{int}}}{2}
		-\ii \frac{\beta_2}{2}\,\partial_\tau^2
		+\ii \gamma \lvert E^{(n)}(z,\tau)\rvert^2
		\right) E^{(n)}(z,\tau),\\
		\label{eq:bd}
		E^{(n)}(0,\tau)&=\sqrt{\rho_{\mathrm{in}}}\,E^{(n-1)}(L,\tau)\,\e^{\ii\delta}
		+ \sqrt{\theta_{\mathrm{in}}P_{\mathrm{in}}}\,\e^{\ii\varphi_n}
	\end{align}
\end{subequations}
consists of a nonlinear Schrödinger equation (NLSE) Eq.~\eqref{eq:nlse}, which gives the evolution of the complex field envelope $E^{(n)}(z, \tau)$ at roundtrip $n$ along the propagation coordinate $z\in [0,L]$ with $L$ the cavity length, and the \textit{fast-time} $\tau$ denotes the time in a reference frame co-moving with the field at the group velocity. The boundary condition Eq.~\eqref{eq:bd}, connects the field envelope at successive roundtrips. The cavity is driven by an input field of power $P_{\mathrm{in}}$ at angular frequency $\omega_0$, phase modulated by the input $\varphi_n$. Here, $\alpha_{\mathrm{int}}$ is the propagation loss; 
$\beta_2 = \left.\mathrm{d}^2\beta/\mathrm{d}\omega^2\right|_{\omega=\omega_0}$ is the second-order dispersion coefficient; 
$\gamma$ is the Kerr nonlinearity coefficient of the fiber; 
$\theta_{\mathrm{in}}$ is the input-coupler reflection coefficient and $\rho_{\mathrm{in}}=1-\theta_{\mathrm{in}}$ is the corresponding transmission coefficient; 
and $\delta$ is the net phase detuning between the driving laser and the nearest cavity resonance.

\subsection{Lugiato--Lefever equation} 
The Lugiato--Lefever Equation (LLE) is a mean-field approximation that captures the dynamics of the intracavity field roundtrip after roundtrip. It is derived from the more general Ikeda map Eq.~\eqref{eq:ikeda_map} when: (i) the field envelope only changes slightly from one roundtrip to the next, and (ii) the total accumulated phase shift per roundtrip is small, i.e., \(\lvert \delta - \gamma L \lvert E\rvert^2 \rvert \ll 1\). The intracavity field dynamics can then be approximated by a single continuous-time partial differential equation~\cite{Haelterman1992}.
\begin{equation}
	t_R\partial_t E(t,\tau) = 
	\left(-\frac{\Lambda}{2} - \ii\delta - \ii\frac{\beta_2 L}{2} \partial^2_{\tau} + \ii\gamma L |E(t,\tau)|^2\right) E(t,\tau) + \sqrt{\theta_{\mathrm{in}}P_{\mathrm{in}}}\, \e^{\ii \varphi(t)},
	\label{eq:LLE}
\end{equation}
where $t_R\,=\,L/v_g$ is the cavity roundtrip time, and $v_g$ is the group velocity of the optical field. The \textit{slow-time} is defined as $t\,=\,n\,t_R$, where $n$ is the roundtrip index, $\Lambda\,=\,\theta_{\mathrm{in}} + \alpha_{\mathrm{int}} L$ is the total cavity loss per roundtrip, and the continuous drive phase $\varphi(t)$ is defined as $\varphi(t)\,=\,\varphi_n$ for $t\,\in\,[n t_R,(n+1)t_R)$. 

With anomalous dispersion, the LLE model (and similarly the Ikeda map) admits three classes of stationary states: homogeneous steady states, patterned states arising from modulational instability (MI), and localized dissipative cavity solitons that coexist with a stable CW background. The latter exists only within a bounded region of the pump–detuning parameter space~\cite{coen2013}.

\subsection{Reduced Model}
The description of cavity solitons can be further simplified by adopting a Lagrangian approach. We take as our starting point the LLE Eq.~\eqref{eq:LLE}, in its Lagrangian formulation. Because slow-time phase perturbations of the input pump do not induce frequency shifts, and because the soliton does not drift in time, by neglecting the background on top of which the soliton sits, one can take as ansatz $E_{\mathrm{CS}}(t,\tau)\,=\,\eta(t)\,\mathrm{sech}\!\left(\frac{\eta(t)\tau}{\sqrt{-\beta_2/\gamma}}\right)\exp\!\left[\ii\big(\phi(t)+\varphi(t)\big)\right]$, where $\eta(t)$ is the soliton amplitude, $\varphi(t)$ denotes the imposed phase modulation of the driving field, and $\phi(t)$ is the soliton phase relative to the instantaneous drive phase. The corresponding Euler–Lagrange equations governing the dynamics of the soliton parameters read (see Appendix):
\begin{subequations}
	\begin{align}
		t_R \dot{\eta}(t) &= -\Lambda\,\eta(t) + \pi\sqrt{\theta_{\mathrm{in}}P_{\mathrm{in}}} \cos\,\bigl(\phi(t)\bigr), \\
		t_R \dot{\phi}(t) &= -\delta - t_R\dot{\varphi}(t) + \frac{1}{2} L\gamma\,\eta^2(t).
	\end{align}
	\label{eq:EOMs}
\end{subequations}
When \(\dot{\varphi}(t)\,=\,0\), the stationary solution of Eqs.~\eqref{eq:EOMs} admits soliton amplitude \(\eta^*\,=\,\sqrt{\frac{2 \delta }{L \gamma}}\) and phase \(\cos\left(\phi^*\right)\,=\,\frac{ \Lambda \eta^*}{\pi \sqrt{\theta_{\mathrm{in}}\,P_{\mathrm{in}}}}\). The latter equation also determines the maximum soliton detuning allowed by the system parameters, given by \(\delta_{\max}\,=\,\frac{\pi^2 \gamma \theta_{\mathrm{in}} P_{\mathrm{in}} L}{2\Lambda^2}\). Fig.~\ref{fig:Losses} shows the stability region as a function of the input phase standard deviation $\sigma_{\varphi}$ for different cavity losses $\Lambda$. Lower-loss cavities exhibit a broader stability range, in good agreement with the prediction for $\delta_{\max}$.

Eqs.~\eqref{eq:EOMs} also provide a semi-analytic model of how the soliton reservoir computer works. Consider a time-dependent input phase $\varphi(t)$. Upon solving Eqs.~\eqref{eq:EOMs} one obtains the time-dependent soliton parameters $(\eta(t),\phi(t))$. Taking the Fourier transform of the resulting $E_{\mathrm{CS}} (t,\tau)$ with respect to $\tau$ and squaring gives the spectral power density of the soliton $\tilde{P}_{CS}(t,\omega)$:
\begin{equation}
	\tilde{P}_{CS}(t,\omega) = \frac{\pi^2 \vert \beta_2 \vert}{\gamma}
	\mathrm{sech}^2\!\left( \frac{\pi\sqrt{-\beta_2/\gamma}}{2\eta(t)}\,\omega \right).
	\label{Eq:Semianalytic}
\end{equation}
The different output nodes of the reservoir are the powers in the spectral slices illustrated in Fig.~\ref{fig:specbreath}. These depend nonlinearly on $\eta(t)$, which in turn depends nonlinearly on the input phase $\varphi(t)$.

We note that the relaxation oscillations observed in Fig.~\ref{fig:relax} can also be obtained from Eqs.~\eqref{eq:EOMs}. To this end, one can linearize Eqs.~\eqref{eq:EOMs} around the stationary soliton solution $(\eta^*, \phi^*)$ yielding the equations of a damped harmonic oscillator with a damping rate determined by the photon lifetime \(t_{\mathrm{ph}}\,=\,t_R/\Lambda\).

\section{Numerical Simulation}
\label{sec:NS}
\begin{figure}[htbp]
	\centering
	\begin{subfigure}[b]{0.4\textwidth}
		\centering
		\vspace{0pt}
		\includegraphics[width=\textwidth]{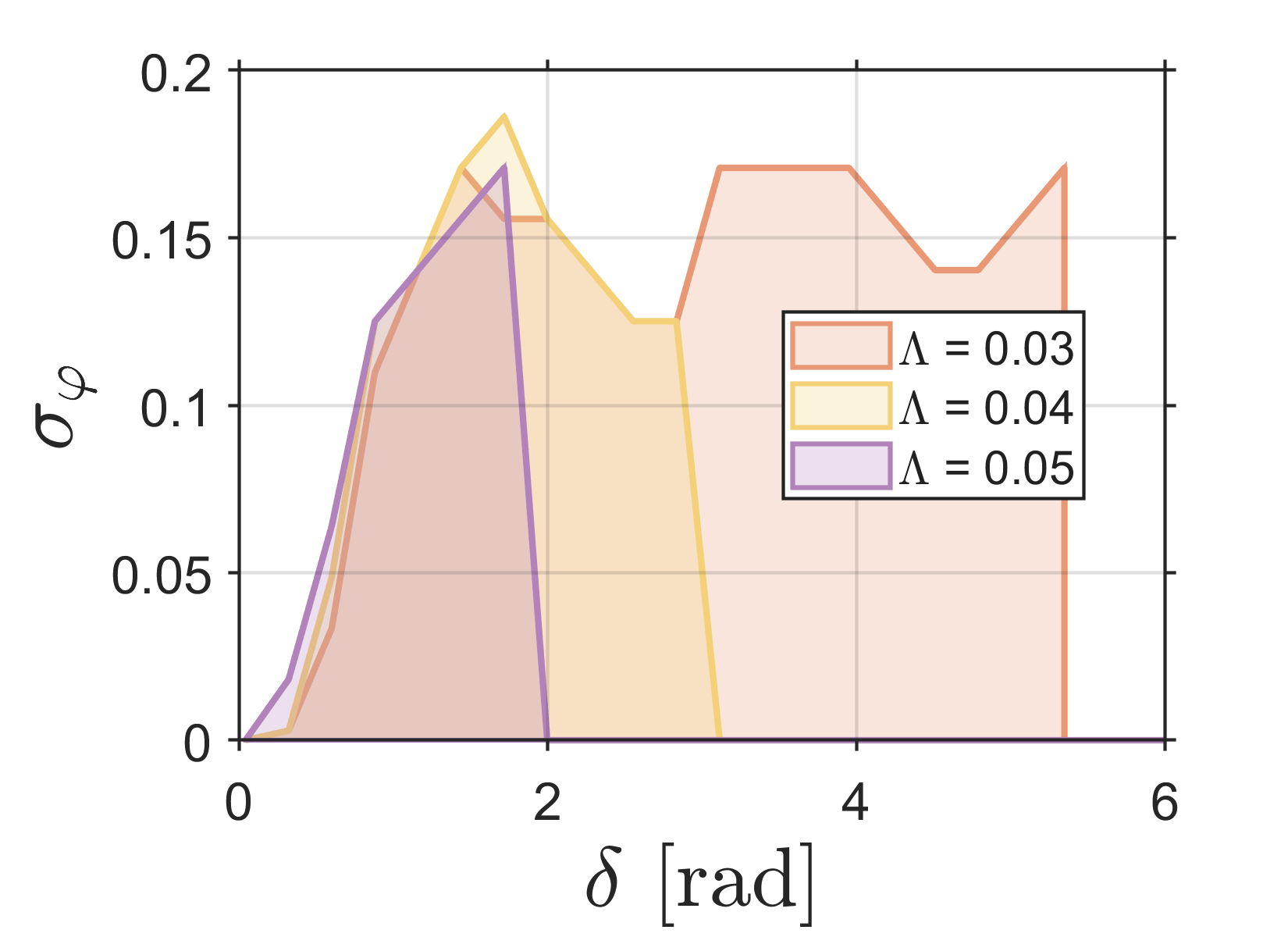}
		\caption{Operating regimes vs. $\Lambda$ and $\delta$}
		\label{fig:Losses}
	\end{subfigure}
	\hspace{.3cm}
	\begin{subfigure}[b]{0.4\textwidth}
		\centering
		\vspace{0pt}
		\includegraphics[width=\textwidth]{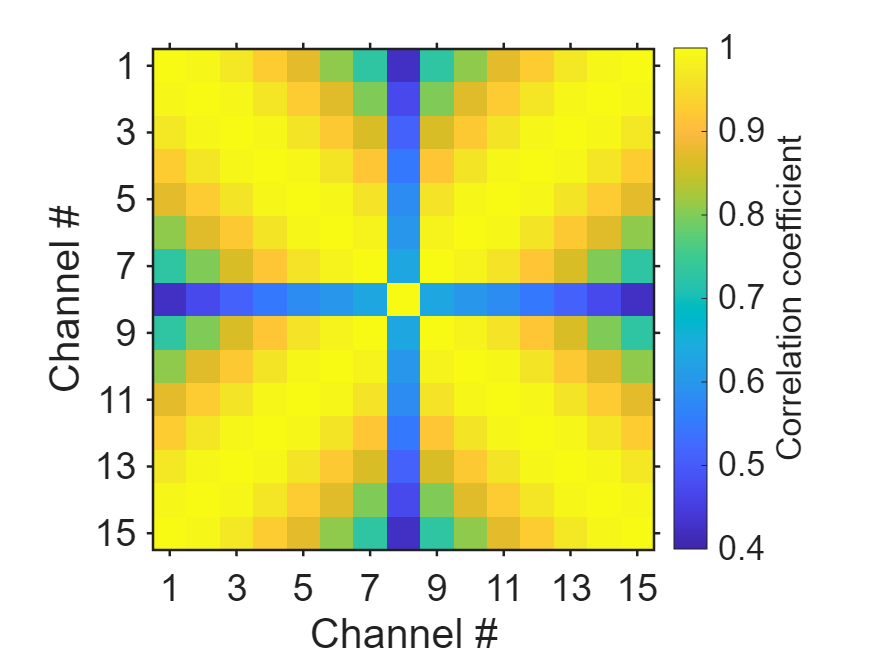}
		\caption{Correlation coefficient of frequency channels}
		\label{fig:corecoef}
	\end{subfigure}
	\caption{Analysis of cavity solitons under phase modulation of the driving field. (a) Stable operational regimes as a function of phase standard deviation $\sigma_{\varphi}$ and detuning $\delta$ for $\Lambda \in \{0.03, 0.04, 0.05\}$. (b) Correlation coefficient corresponding to Fig.~\ref{fig:specbreath}. The above simulations were generated using the Ikeda map model, with other parameters the same as in Fig.~\ref{fig:mainpanel}.}
	\label{fig:OverallFigure}
\end{figure}

Numerical simulations allow us to explore parameter regimes that are not easily accessible in the current experimental realization, to assess system performance in the absence of experimental noise, and to identify the physical mechanisms underlying the reservoir performance. Fig.~\ref{fig:corecoef} shows the correlation matrix of the different spectral channels associated with the dynamics in Fig.~\ref{fig:specbreath}. This matrix reveals that channels closer to the pump wavelength, where the homogeneous continuous-wave field constitutes a distinct steady-state solution of the system, exhibit d correlation with the rest of the spectrum. Conversely, channels with larger frequency offset with respect to the pump are more strongly correlated, as their dynamics are predominantly governed by the cavity soliton. Notably, the soliton spectrum exhibits a symmetric profile, implying that nodes at equidistant positive and negative frequency shift provide redundant dynamics. To break the spectral symmetry and reduce redundancy among reservoir nodes, we use \(N\) bandpass filters with fixed bandwidth, whose centers are randomly distributed across the soliton spectrum to define the reservoir nodes. This random configuration is fixed using MATLAB's \texttt{rng} function across all simulations reported below, ensuring reproducibility.

For numerical integration, the LLE Eq.~\eqref{eq:LLE} and the NLSE Eq.~\eqref{eq:nlse} were solved using the split-step Fourier method~\cite{argwal}, while the reduced model Eq.~\eqref{eq:EOMs} was integrated via a fourth-order Runge--Kutta scheme. Unless explicitly stated, the numerical simulations described below use the Ikeda map to model a passive cavity without added noise. The parameters used closely match the experimental system: $\beta_2\,=\,-23\,\mathrm{ps}^2\mathrm{km}^{-1}$, $\gamma\,=\,1.3\,\mathrm{W}^{-1}\mathrm{km}^{-1}$, $L\,=\,50\,\mathrm{m}$, $\theta_{\mathrm{in}}\,=\,10\,\%$.

We note that in the experimental setup (see Section~\ref{sec:EXP_setup}), a low-gain amplifier is inserted in the cavity. This reduces the effective loss to mimic a low-loss cavity, which lowers the required driving power to sustain the cavity solitons. However, as a result of the small gain saturation by the background, the effective roundtrip loss $\Lambda_{\e}$ of the active cavity depends on the cavity detuning~\cite{Englebert2021}. Numerical simulations show that, for detunings $\delta \gtrsim 0.5~\mathrm{rad}$, this effective loss remains approximately constant, allowing the active cavity to be mapped onto an equivalent passive cavity with fixed roundtrip loss. Throughout our simulations, we use a total cavity loss of $\Lambda\,=\,0.03$ (dropping the subscript ${e}$ to denote the passive cavity approximation). The corresponding field transmission $\sqrt{1-\Lambda}$ enters the Ikeda-map boundary condition in Eq.~\eqref{eq:bd}. Other parameters, such as input power $P_{\mathrm{in}}$, cavity detuning $\delta$, and the number of neurons $N$, are specified for each task.

\subsection{Beyond the mean-field model: Kelly waves and information processing}
\label{sec:modelCompare}
\begin{figure}[htbp]
	\centering
	\begin{subfigure}[b]{0.24\textwidth}
		\centering
		\includegraphics[width=\textwidth]{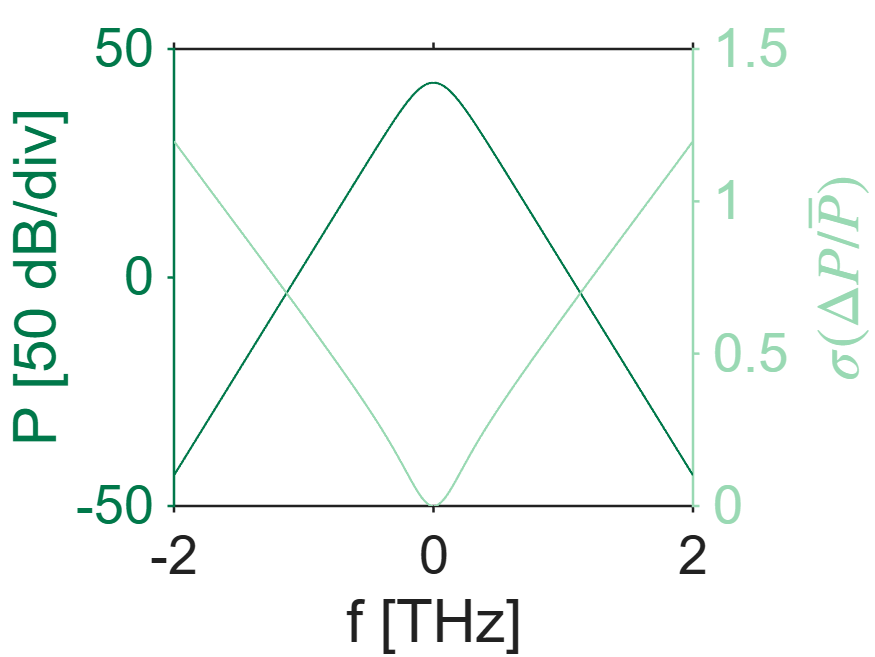}
		\caption{Reduced Model}
		\label{fig:reduced_evol}
	\end{subfigure}
	\hfill
	\begin{subfigure}[b]{0.24\textwidth}
		\centering
		\includegraphics[width=\textwidth]{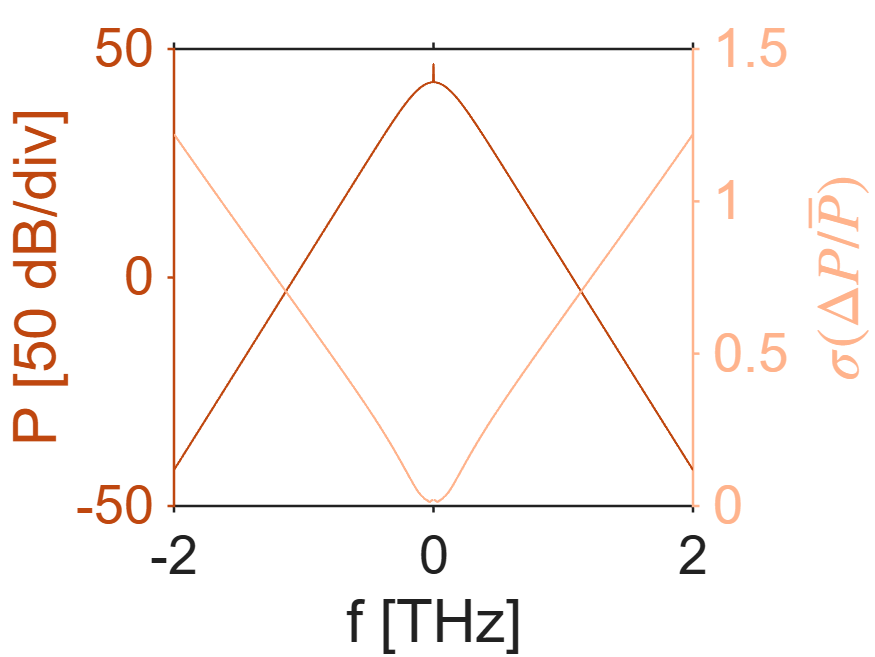}
		\caption{LLE Model}
		\label{fig:LLE_evol}
	\end{subfigure}
	\hfill
	\begin{subfigure}[b]{0.24\textwidth}
		\centering
		\includegraphics[width=\textwidth]{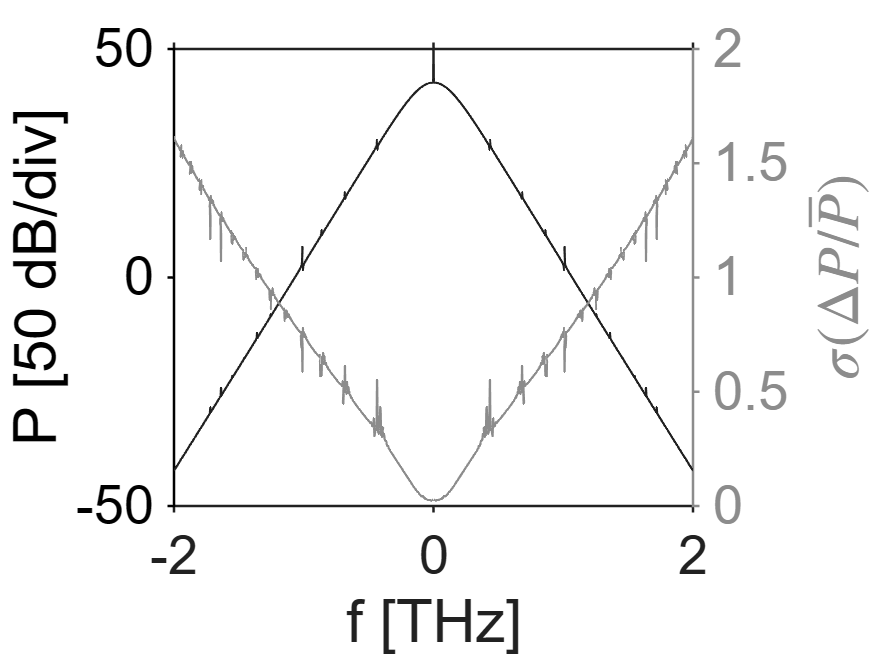}
		\caption{Ikeda Map}
		\label{fig:Ikeda1_evol}
	\end{subfigure}
	\hfill
	\begin{subfigure}[b]{0.24\textwidth}
		\centering
		\includegraphics[width=\textwidth]{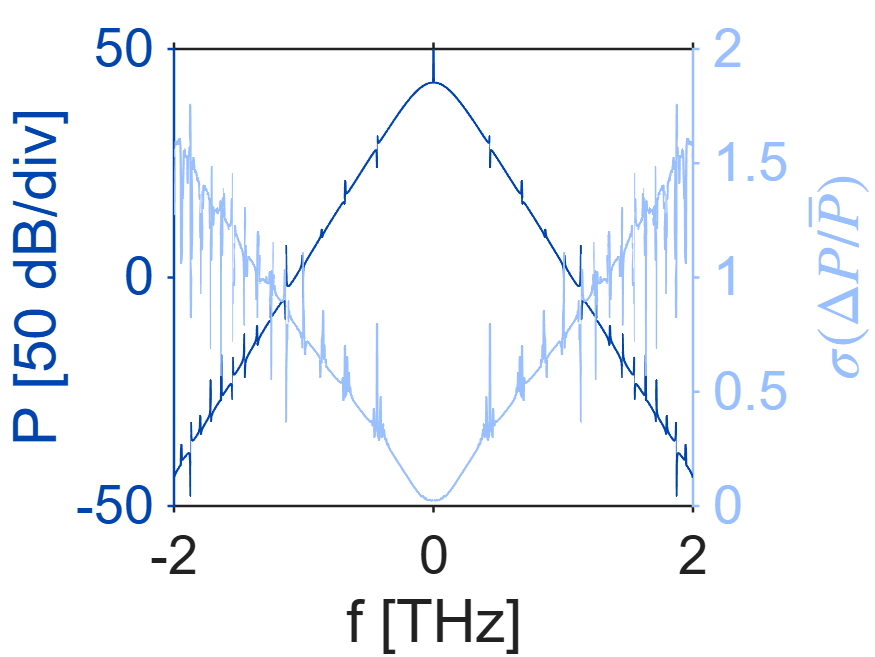}
		\caption{Kelly Waves}
		\label{fig:Ikeda2_evol}
	\end{subfigure}
	\vfill
	\begin{subfigure}[b]{0.8\textwidth}
		\centering
		\includegraphics[width=\textwidth]{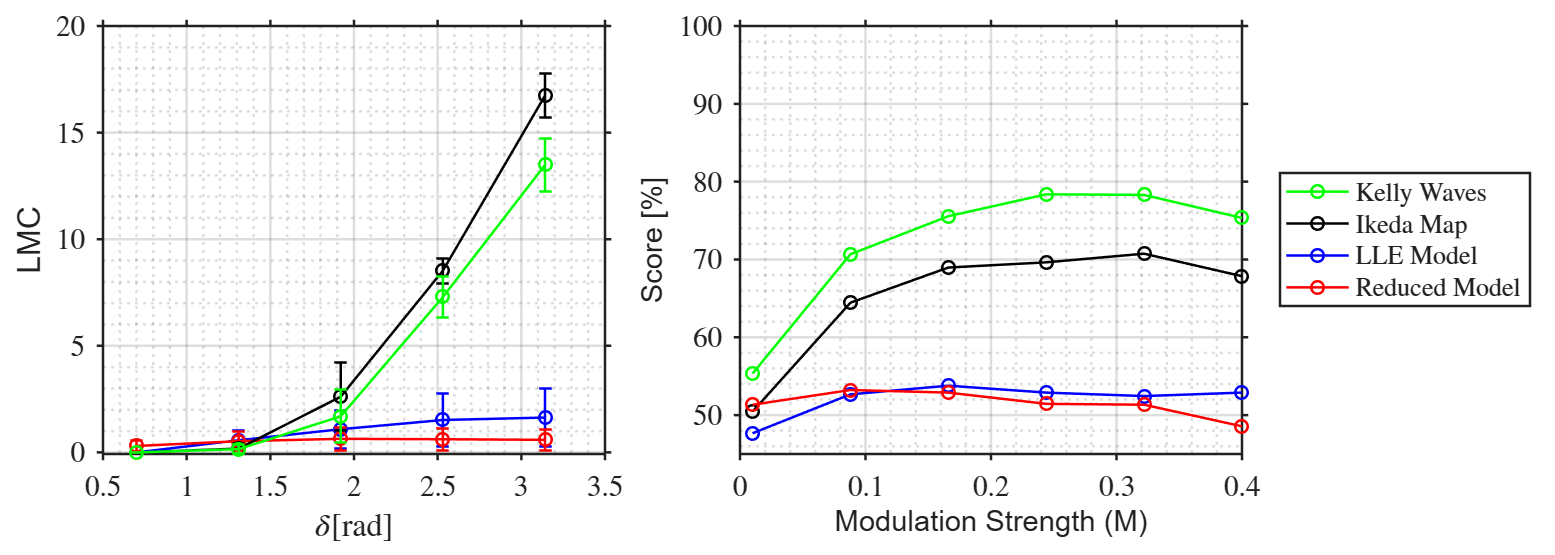}
		\caption{LMC and XOR benchmark for different models}
		\label{fig:Model_benchmarks}
	\end{subfigure}
	\caption{Comparative analysis of different theoretical frameworks for reservoir computing.
		(a--d): Normalized spectral fluctuations $\sigma\!\left(\Delta P/\bar{P}\right)$ and CS spectra versus frequency offset $f$ at $\delta\,=\,2.5\,\mathrm{rad}$. The plot illustrates how different frequency components respond to phase perturbations. Here $\sigma\!\left(\Delta P/\bar{P}\right)$ denotes the standard deviation of the normalized relative power change. 
		(e): Left: Linear Memory Capacity (LMC) as a function of detuning for different models  ($M\,=\,1$). Right: XOR benchmark classification accuracy (Score,\%) as a function of modulation strength $M$, between the current input and 4-step delayed input for parameters: $K\,=\,2000$, $N\,=\,50$, $\delta\,=\,\pi\,\mathrm{rad}$, with filter bandwidth ranges from $96$ to $107\,\mathrm{GHz}$, corresponding to a detuning range of $\delta \in [0.7, \pi]\,\mathrm{rad}$. (All results are obtained for a repetition factor of $q\,=\,10$, and $P_{\text{in}}\,=\,250\,\mathrm{mW}$).}
	\label{fig:Modelcompare}
\end{figure}

In Sec.~\ref{sec:models}, we presented three distinct models describing the dynamical evolution of cavity solitons in response to a phase-encoded driving beam. Here we compare the predictions of the three models. We consider the spectrum and the normalized spectral fluctuations $\frac{\sigma_P}{P}(f)$ as a function of constant frequency offset from the pump beam $f$ to illustrate how different frequency components respond to the phase perturbation. To illustrate how these differences alter the reservoir dynamics, we evaluate performance on two benchmark tasks: Linear Memory Capacity (LMC) and the XOR operation between the current input and 4-step delayed input, with CS spectrum truncated at $-50\mathrm{dB}$ (Fig.~\ref{fig:Modelcompare}, upper panel).

The LMC benchmark evaluates the reservoir’s ability to recall past inputs. Inputs were sampled $\mathrm{i.i.d.}$ from the uniform distribution over \([0,1]\) with modulation strength $M$ corresponding to \(\sigma_\varphi\,=\,M/\sqrt{12}\). For each delay $d$, we train the system to reconstruct the $d$-step delayed input $\mathbf{u}_{\mathrm{train}}^{[d]}$. Performance is quantified by the coefficient of determination on the test set
\[
R_d^2 = 1 - \frac{\lVert \mathbf{y}_{\mathrm{test}} - \mathbf{u}_{\mathrm{test}}^{[d]} \rVert_2^2}
{\lVert \mathbf{u}_{\mathrm{test}}^{[d]} - \overline{\mathbf{u}}_{\mathrm{test}}^{[d]} \rVert_2^2},
\]
where $\mathbf{y}_{\mathrm{test}}$ denotes the reservoir's prediction, $\mathbf{u}_{\mathrm{test}}^{[d]}$ is the corresponding delayed target, and $\mathbf{\bar{u}}_{\mathrm{test}}^{[d]}$ is its mean. The total LMC is obtained by summing over all these coefficients, which is theoretically upper-bounded by the number of neurons in the reservoir~\cite{Yeager_shortterm}. We used a 70/30 train/test split and a maximum delay equal to the total number of neurons $N$. 

The XOR task consists of providing as input a sequence of binary symbols $\{0, 1\}$ and computing the XOR of two symbols separated by a fixed delay. As the XOR operation is not linearly separable, a purely linear reservoir will perform no better than random guessing, leading to a misclassification rate of 50\%. Consequently the XOR task measures both the memory and nonlinear information processing capabilities of a reservoir computing system.

\subsubsection{Reduced and mean-field descriptions of cavity solitons}
The reduced model is the simplest. It is useful because it allows analytical predictions, see Eq.~\eqref{Eq:Semianalytic}. The spectrum and the normalized spectral fluctuations $\frac{\sigma_P}{P}$ obtained using the reduced model are plotted in Fig.~\ref{fig:reduced_evol}. We see that for large frequency shift $\frac{\sigma_P}{P}$ is linear in $\vert f\vert$. This is predicted by the reduced model. Indeed, assuming that $f$ is sufficiently large that we can approximate Eq.~\eqref{Eq:Semianalytic} by a decaying exponential, writing $\eta(t)\,=\,\eta^*+\Delta\eta(t)$ with $\eta^*$ the amplitude of the unperturbed soliton, and assuming  $\Delta\eta(t)$ has zero mean and is sufficiently small that we can make a series expansion in $\Delta\eta(t)$, one finds that $\frac{\sigma_P}{P}\,=\,\frac{\pi \sqrt{-\beta_2/\gamma}}{\eta^{*2}}\sigma_\eta \vert f\vert$ is linear in $\vert f \vert$, with $\sigma_\eta$ the standard deviation of $\Delta\eta (t)$. 

In Fig.~\ref{fig:Model_benchmarks} we have plotted the performance of the reduced model on the two tasks. We observe that this approach yields rather low performance. This can be explained by Eq.~\eqref{Eq:Semianalytic}, which shows that the power spectral density depends only on the soliton amplitude $\eta$, while phase information $\phi$ is lost, reducing the dimensionality of the system's state space, limiting its computational capacity. The Lugiato--Lefever equation is more realistic than the reduced model as it explicitly incorporates the background field (omitted in the Lagrangian analysis). However, the spectrum and the spectral fluctuations $\frac{\sigma_P}{P}$ plotted in Fig.~\ref{fig:LLE_evol} are very similar to the reduced model except for a small difference close to $f\,=\,0$, where the spectral interference between soliton and background beam takes place. The performance on tasks reported in Fig.~\ref{fig:Model_benchmarks} is also very close to the performance of the reduced model.

\subsubsection{Ikeda map and contribution of Kelly waves} 
As shown in Fig.~\ref{fig:Model_benchmarks}, the Ikeda map gives significantly better results than the reduced and the LLE models on benchmark tasks. We attribute this to the fact that the Ikeda map captures additional dynamical features absent in the reduced and the LLE models~\cite{Hansson2015}.

The difference between the models is evident in the spectrum shown in Fig.~\ref{fig:Ikeda1_evol}, which reveals the presence of Kelly sidebands. The plot of $\frac{\sigma_P}{P}$ shows strong peaks at the frequencies of the Kelly sidebands, indicating that these spectral components respond to phase perturbations. This interpretation is consistent with the dependence of the system performance on the cavity detuning $\delta$ reported in Fig.~\ref{fig:Model_benchmarks}. As $\delta$ increases, the soliton becomes temporally narrower and spectrally broader, leading to the appearance of additional Kelly sidebands in the spectrum. 

To further investigate the role of the Kelly sidebands, we considered in Fig.~\ref {fig:Ikeda2_evol} the Ikeda map with a longitudinally varying dispersion profile $\beta_2(z)$. The average second-order dispersion is the same as in Fig.~\ref{fig:Ikeda1_evol} ($\beta_2\,=\,-23\,\mathrm{ps}^2\mathrm{km}^{-1}$), but is taken to be $-18\,\mathrm{ps}^2\mathrm{km}^{-1}$ over a 1-meter segment inside the cavity (e.g., between 13 and 14 meters). This gives rise to enhanced Kelly sidebands (their locations are identical, but their amplitude is larger), and larger peaks in $\frac{\sigma_P}{P}$ as shown in Fig.~\ref{fig:Ikeda2_evol}. Figure.~\ref{fig:Model_benchmarks} shows that enhancing the Kelly sidebands modifies performance, but in a way that is task-dependent, confirming that they indeed influence the reservoir performance.

\subsection{Reservoir Computing Benchmark}
\label{sec:RCbench}

\begin{figure}[htbp]
	\centering
	\begin{subfigure}[b]{0.35\textwidth}
		\centering
		\vspace{0pt}
		\includegraphics[width=\textwidth]{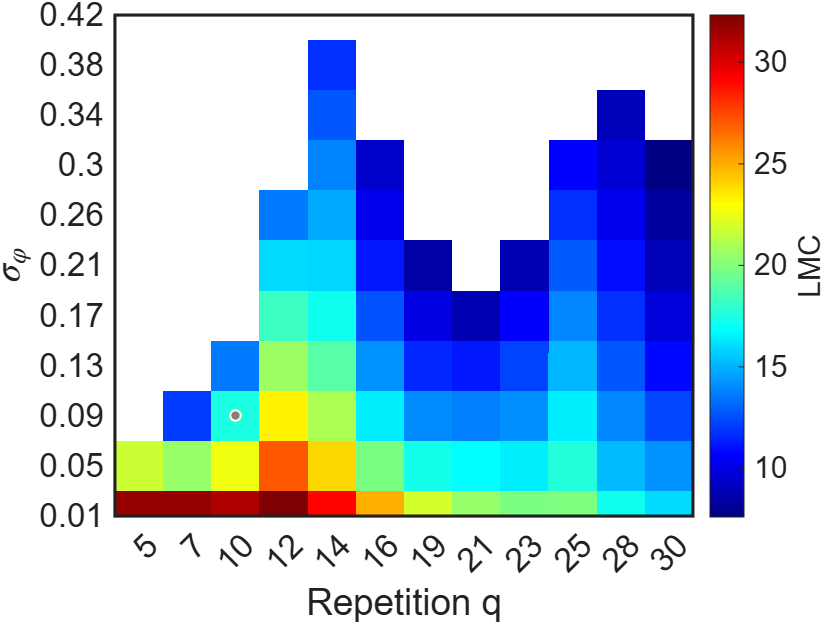}
		\caption{LMC versus $\sigma_{\varphi}$ and $q$}
		\label{fig:LMC-Mq}
	\end{subfigure}
	\hspace{.5cm}
	\begin{subfigure}[b]{0.35\textwidth}
		\centering
		\vspace{0pt}
		\includegraphics[width=\linewidth]{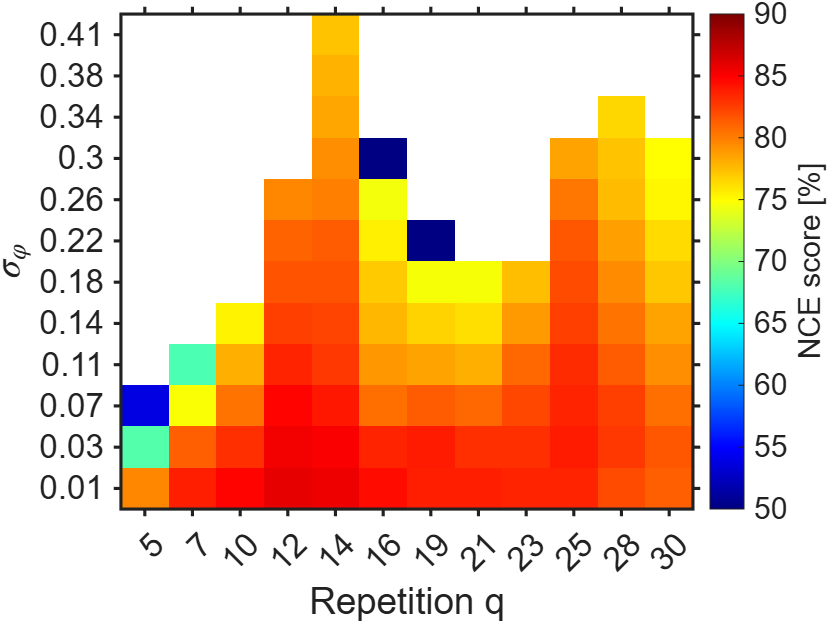}
		\caption{NCE versus $\sigma_{\varphi}$ and $q$}
		\label{fig:NLCE}
	\end{subfigure}
	\vspace{0.2cm}
	\begin{subfigure}[b]{\textwidth}
		\centering
		\vspace{0pt}
		\includegraphics[width=0.85\linewidth]{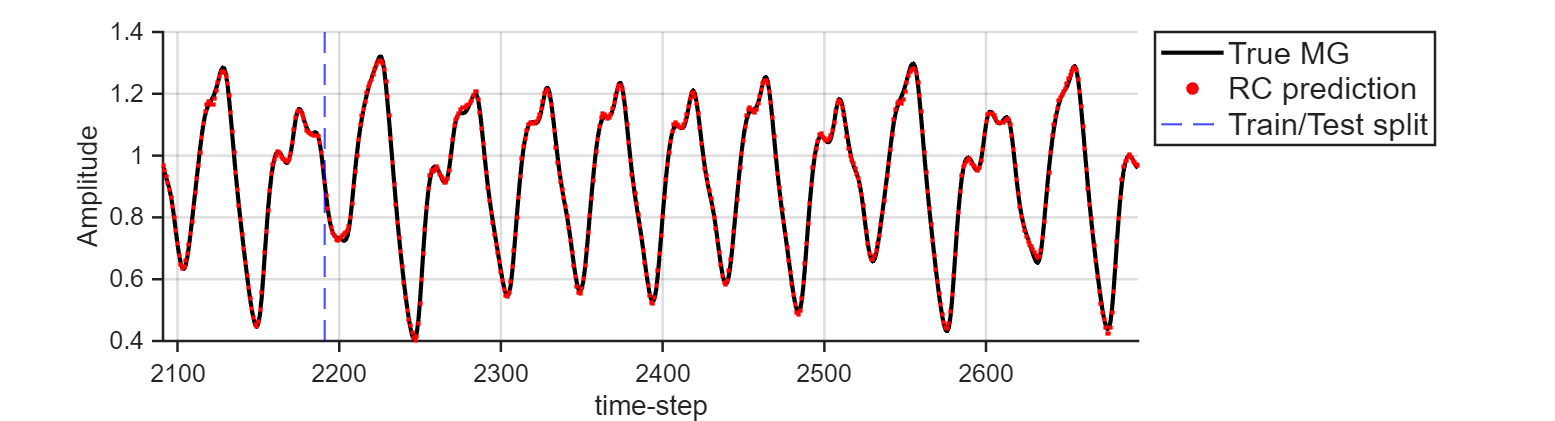}
		\caption{Mackey–Glass System}
		\label{fig:MGsystem}
	\end{subfigure}
	\caption{Reservoir computing benchmarks. Panels (a--b) show performance over the parameter space defined by the input-phase standard deviation $\sigma_{\varphi}$ and repetition factor $q$, while (c) shows the performance at a single coordinate within this space.
		(a) Linear Memory Capacity (LMC) ($K=5000$).
		(b) Nonlinear Channel Equalization (NCE) at channel SNR = 12~dB ($K=7000$).
		(c) Six-step-ahead prediction of the Mackey-Glass system at $\sigma_{\varphi} = 0.24$, $q=10$, and $K=3000$. The target data (black line), reservoir prediction (red dots), and train-test split (vertical blue dashed line) are shown.
		Parameters: cavity detuning $\delta=2.5~\mathrm{rad}$ and input power $P_{\mathrm{in}}=200~\mathrm{mW}$.}
	\label{fig:M_Q}
\end{figure} 
Here, we evaluate the system performance using standard reservoir computing benchmark tasks, using the Ikeda map as a dynamical model. We truncate the CS spectrum at $-100\,\mathrm{dB}$, corresponding to a span of $7.2\,\mathrm{THz}$. This span is divided into $N\,=\,50$ frequency bands, each with a bandwidth of $145\,\mathrm{GHz}$. This wide span ensures that many Kelly waves are included in the information processing. The training was performed on the reservoir states using ridge regression Eq.~\eqref{eq:training}, and the error was reported on the test data according to Eq.~\eqref{eq:NRMSE}. 

\subsubsection{Linear Memory Capacity}
Figure~\ref{fig:LMC-Mq} shows the total LMC as a function of the input-phase standard deviation $\sigma_{\varphi}$ and repetition factor $q$. The highest LMC of 32 was obtained at very low modulation strength corresponding to $\sigma_{\varphi}\,=\,0.01$ and small repetition factor $q\,=\,5$. This configuration corresponds to a very linear regime of operation. The white regions in Fig.~\ref{fig:LMC-Mq} correspond to parameter regimes where the soliton collapses, caused by high modulation strength. This threshold modulation strength depends on $q$, presumably because there is an interplay between the frequency of the relaxation oscillations (see Fig.~\ref{fig:relax}) and the repetition factor $q$.

\subsubsection{Nonlinear Channel Equalization}
We benchmarked the cavity-soliton reservoir computer on the Nonlinear Channel Equalization (NCE) task used in \cite{jaeger2004}. The goal is to reconstruct the transmitted symbols $d(n)$ after propagation through a noisy, nonlinearly distorted channel producing $r(n)$:
\begin{subequations}
	\begin{align}
		z(n) &= 0.08\,d(n+2) - 0.12\,d(n+1) + d(n) + 0.18\,d(n-1) - 0.1\,d(n-2) \\
		& \quad + 0.091\,d(n-3) - 0.05\,d(n-4) + 0.04\,d(n-5) + 0.03\,d(n-6) + 0.01\,d(n-7), \notag \\
		r(n) &= z(n) + 0.036\,z^2(n) - 0.011\,z^3(n) + \nu(n),
	\end{align}
\end{subequations}
where $\nu(n)$ is Gaussian noise with zero mean and variance set by the channel Signal-to-Noise Ratio (SNR). The transmitted symbols $d(n)$ are randomly drawn from four values $\{-3,-1,1,3\}$, with the channel SNR set at $12\,\mathrm{dB}$. The performance is measured by the classification accuracy (i.e., $1$ minus the Symbol Error Rate). Figure~\ref{fig:NLCE} shows the classification accuracy of the soliton reservoir as a function of the standard deviation of inputs $\sigma_{\varphi}$ and the repetition factor $q$. The best performance of $85.8\%$ accuracy was obtained for $\sigma_{\varphi}\,=\,0.01$ and a repetition factor of $q\,=\,12$. The classification accuracy could be further improved by increasing the input power to $P_{\mathrm{in}}\,=\,500\,\mathrm{mW}$, which expands the stability region of the system. In this configuration, we achieved a peak classification accuracy of $90.5\%$ at $\sigma_{\varphi}\,=\,0.11$ and a repetition factor of $q\,=\,10$. This performance surpasses that of the linear reservoir, which has a maximum accuracy of $87\%$ for the same channel SNR, and is competitive with the high-performance photonic reservoir reported in~\cite{Vinckier2015}, which reported a classification accuracy of $90\%$ given the same channel SNR.

\subsubsection{Mackey–Glass System}
The Mackey--Glass (MG) system~\cite{MGsystem} is a delayed differential equation 
\begin{equation}
	\dot{x}(t) = \frac{0.2\,x(t-\tau)}{1 + x^{10}(t-\tau)} - 0.1\,x(t),
	\label{Eq:MG}
\end{equation}
with delay parameter \(\tau\,=\,17\). We solved Eq.~\eqref{Eq:MG} using the Euler integration method with time step \(\Delta t\,=\,1\). We assessed the ability of the soliton reservoir, given the past of the trajectory, to predict its future $6$, $10$ and $15$ steps ahead. For the 6-step ahead prediction, with \(\sigma_{\varphi}\,=\,0.24\) and \(q\,=\,10\) we obtained a NRMSE of 0.01 (see Fig.~\ref{fig:MGsystem}). Performance was found to degrade with larger time steps, with NRMSE values of 0.02 and 0.03 when predicting 10 and 15 time steps ahead respectively. This performance is significantly better than a linear reservoir, which yields a NRMSE of 0.14 for six-step-ahead prediction under the same numerical conditions, and competitive with a recent all-optical echo-state network~\cite{Kaushik2025}, which reported an NRMSE of 0.06 for a 5-step-ahead prediction.

\section{Experimental Setup}
\label{sec:EXP_setup}

\begin{figure}[htbp]
	\centering
	\begin{subfigure}[b]{\textwidth}
		\centering
		\vspace{0pt}
		\includegraphics[width=0.9\textwidth]{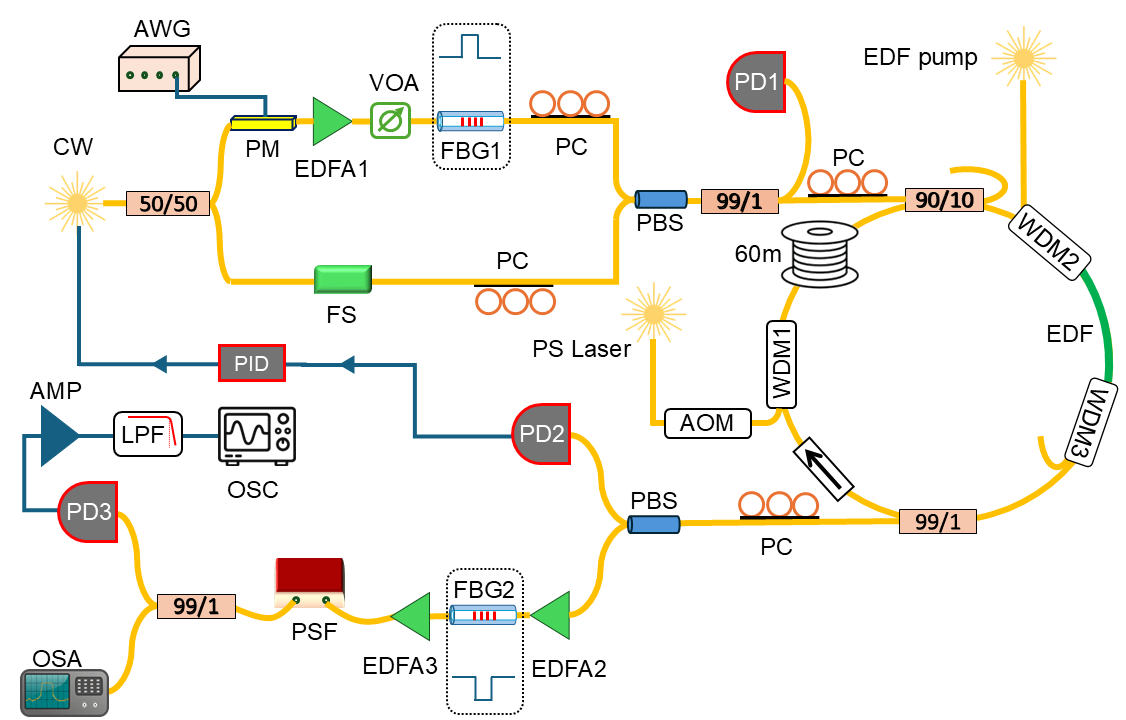}
		\caption{Experimental setup for frequency-multiplexed reservoir computing using cavity solitons}
		\label{fig:expsetup}
	\end{subfigure}
	\vfill
	\begin{subfigure}[b]{0.4\textwidth}
		\centering
		\vspace{0pt}
		\includegraphics[width=\linewidth]{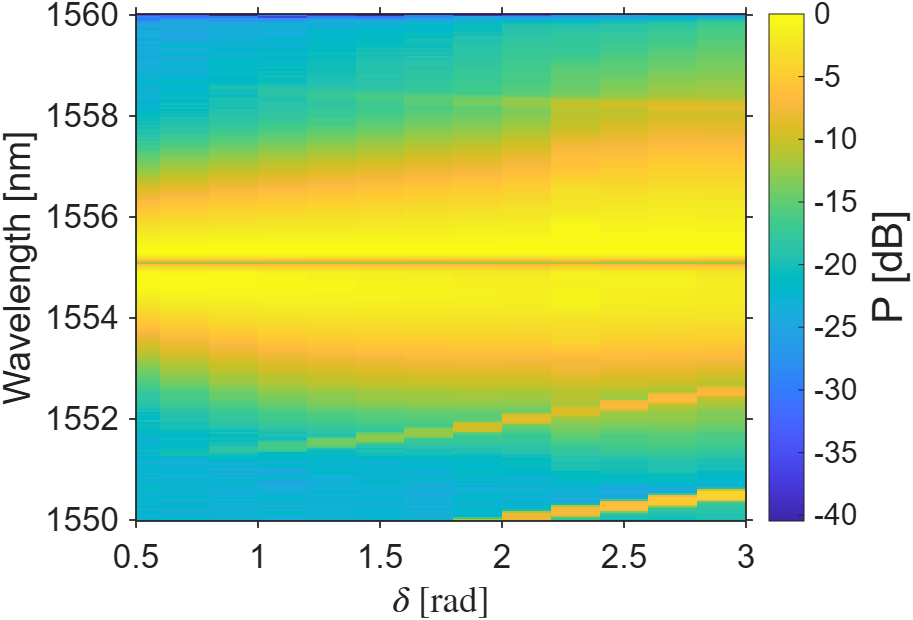}
		\caption{Spectral profile of cavity solitons versus detuning while the background has been removed}
		\label{fig:expsetup_spec}
	\end{subfigure}
	\hspace{.5cm}
	\begin{subfigure}[b]{0.45\textwidth}
		\centering
		\vspace{0pt}
		\includegraphics[width=\linewidth]{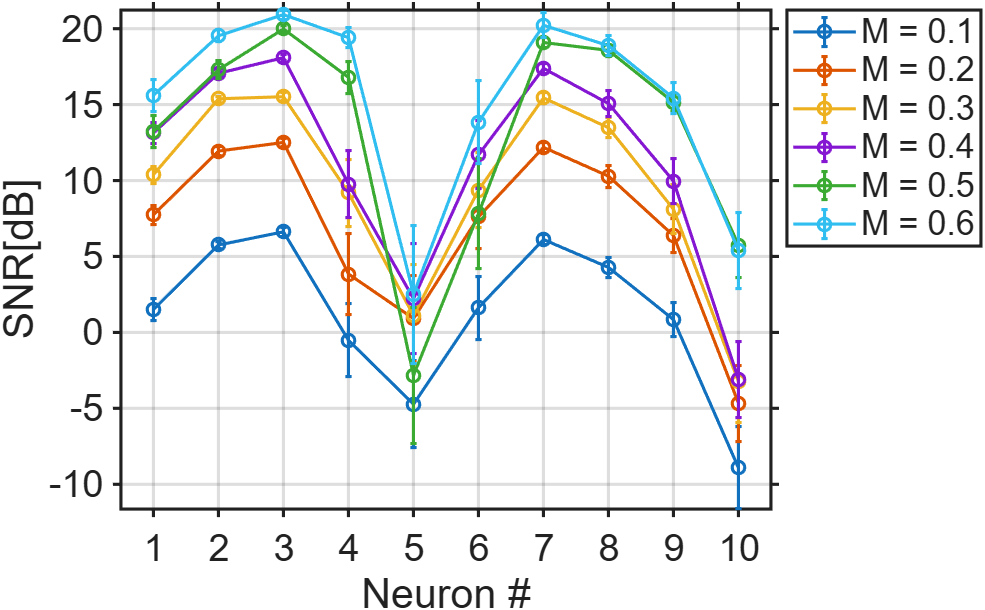}
		\caption{Node-wise SNR for different values of modulation strength $M$ at $\delta\,=\,1.5\,\mathrm{rad}$}
		\label{fig:snrvsM}
	\end{subfigure}
	\caption{(a) Experimental setup. A continuous-wave laser (CW) is split using a 50:50 coupler. In the upper arm, an arbitrary waveform generator (AWG) encodes information via a phase modulator (PM). The modulated signal is amplified by an erbium-doped fiber amplifier (EDFA1), trimmed using a variable optical attenuator (VOA), and spectrally cleaned by a fiber Bragg grating (FBG1). In the lower arm, a frequency shifter (FS) generates a control signal, with a power of approximately $10\ \mathrm{ \mu W}$ at the input to the cavity. The two signals are recombined using a polarization beam splitter (PBS). The combined signal is coupled into the cavity via a 90/10 fiber coupler. The cavity consists of a SMF-28 spool, an optical isolator, an erbium-doped fiber (EDF), two 1480/1550\,$\mathrm{nm}$ wavelength-division multiplexers (WDM2 and WDM3), and one 1535/1550\,$\mathrm{nm}$ WDM (WDM1). One percent of the intracavity field is tapped. Stabilization of the detuning $\delta$ is implemented using a proportional-integral-derivative (PID) controller. After removing the CW background using FBG2 and re-amplifying the signal using EDFA2 and EDFA3, a programmable spectral filter (PSF) applies bandpass filters across the output soliton spectrum. One percent of the filtered output is sent to an optical spectrum analyzer (OSA) for monitoring, while 99\% of the filtered output is converted to an electrical signal by a photodiode (PD3) followed by a $10\,\mathrm{MHz}$ low-pass filter (LPF), amplified electrically (Amp), and recorded by the oscilloscope (OSC). PC: polarization controller.
		(b) Spectral profile of the cavity solitons as a function of detuning $\delta$. As the detuning increases, the spectrum broadens, and shifts due to Raman self-frequency shift and Kelly sidebands appear. The spectral power for each trace is independently normalized to its maximum power and plotted on a logarithmic scale. 
		(c) The effect of modulation strength, \(M\), on the Signal-to-Noise Ratio (SNR). Here, i.i.d. inputs from the range [0, 1] are randomly sampled and encoded onto the PM. The plot shows the resulting SNR per node, when $M$ is varied from 0.1 to 0.6.}
\end{figure}   

The cavity-soliton reservoir computer setup is depicted in Fig.~\ref{fig:expsetup}. The fiber cavity with free spectral range (FSR) of $3.36\,\mathrm{MHz}$ corresponding to cavity roundtrip time of $t_R\,=\,297\,\mathrm{ns}$ is coherently driven by a CW laser at $1554.94\,\mathrm{nm}$ and consists of a $60\,\mathrm{m}$ fiber spool (SMF-28), an optical isolator to suppress Brillouin lasing, and two couplers (90:10 input and a 99:1 tap coupler for monitoring the cavity dynamics). 

As seen in Fig.~\ref{fig:Losses}, lowering the intracavity losses expands the soliton stability region; our numerical simulations in Fig.~\ref{fig:Model_benchmarks} further demonstrate that operating the reservoir at higher detuning improves performance. Therefore, to compensate for intracavity losses and ensure a high-finesse cavity, $30\,\mathrm{cm}$ of Erbium-Doped Fiber (EDF) is placed inside the cavity and pumped at $2\,\mathrm{W}$ by a Raman laser (Keopsys) via a 1480/1550 $\mathrm{nm}$ WDM (WDM2)~\cite{Englebert2021}. Residual pump power is removed with WDM3. The cavity effective roundtrip loss $\Lambda_{\e}$ is defined as the difference between the total cold-cavity loss $\Lambda$ and the intracavity gain, $\Lambda_{\e}\,=\,\Lambda - g_0L_{\e}$, where $L_{\e}$ and $g_0$ are the length and gain of the EDF. In a regime where the intracavity power is well below the saturation power of the gain medium ($\sim\,1\mathrm{W}$), the effective loss of the cavity was measured to be approximately 3\% corresponding to a cavity finesse of 200. Beyond reducing the effective loss, in our experiments, the gain is not expected to play a role in the cavity dynamics. These values are considerably lower than our measured cold-cavity loss (without gain) of approximately 21\% (finesse $\simeq 26$), and similarly lower than those typical of purely passive fiber cavities, such as the 26\% roundtrip loss reported in Ref.~\cite{Leo2013}. Keeping the parameters identical to those in Fig.~\ref{fig:Losses}, these losses correspond to maximum detunings of $\delta_{\text{max}} = 0.1$ and $0.7$ rad, respectively. Cavity solitons are excited through the cross-phase modulation between the driving field and addressing pulses generated by a mode-locked laser (PriTel FFL) centered at $1535~\mathrm{nm}$ with a repetition rate of $10~\mathrm{MHz}$. The pulses are gated by an AOM and injected into the cavity through a $1535/1550~\mathrm{nm}$ WDM1~\cite{Leo2013}.

To implement reservoir computing with cavity solitons, the input data sequences are generated by an Arbitrary Waveform Generator (AWG) and encoded onto the phase of the CW driving beam using a phase modulator (PM) with half-voltage $V_{\pi}\,=\,4\,\mathrm{V}$. The modulated signal is amplified using a high-power Erbium-Doped Fiber Amplifier (EDFA1). A variable optical attenuator (VOA) is then used to adjust the launch power to the desired value, which is monitored via a 99:1 coupler and PD1. A narrowband $50\,\mathrm{GHz}$ bandpass Fiber Bragg Grating (FBG1-OEWFG-050) filter is inserted to suppress amplified spontaneous emission (ASE) noise originating from EDFA1. At the readout stage, the $1\%$ tapped signal is separated into two orthogonal polarizations using an output PBS. One polarization is used for active cavity detuning stabilization: a CW locking signal generates an error signal that is fed to a proportional–integral–derivative (PID) controller, which adjusts the CW laser frequency. The second polarization is amplified by EDFA2 for subsequent analysis of the soliton dynamics. Given the low duty cycle of the cavity solitons ($\sim\mathrm{ps}$) and the cavity roundtrip time $297\,\mathrm{ns}$, most of the intracavity power and ASE resides in the CW background. Hence, a $50\,\mathrm{GHz}$ bandstop FBG filter (FBG2) is used to filter out the background. The resulting signal is further amplified by EDFA3, and a programmable spectral filter (PSF-Finisar Waveshaper) selects one of $N$ frequency bands from the soliton's spectrum. 

To suppress high-frequency electronic noise while ensuring that the intracavity dynamics over a single roundtrip are fully captured, a $10\,\mathrm{MHz}$ photodiode (NewFocus 2051) with an internal amplifier (Amp) of gain $10^{3}$ is employed (PD3) to measure the optical power in each frequency band. Multiple cavity solitons (typically 10--20) are excited in each RC run to obtain sufficient spectral power at the readout. The soliton density is kept low enough that the solitons remain well separated in time. Since the input encoding rate lies in the kHz range, each reservoir-computing run lasts on the order of milliseconds, which is short enough that the solitons do not reorganize into bound states (e.g., soliton molecules). Under these conditions, the unperturbed solitons remain stable for tens of minutes. The output photodiode (PD3) is further filtered using an external $10~\mathrm{MHz}$ electronic low-pass filter (LPF) before being recorded by the oscilloscope. In this configuration, the reservoir operates in a fully analog manner, while the readout weights and output signals are computed digitally. 

Following Eq.~\eqref{Eq:Semianalytic}, the soliton spectral width scales as $\sqrt{\delta}$.  Fig.~\ref{fig:expsetup_spec} shows that as the cavity detuning is increased, the spectrum broadens and undergoes a Raman self-frequency shift~\cite{Karpov2016}. In the absence of phase perturbations of the driving field and an input power of $250\,\mathrm{mW}$, we observed that solitons persist in the cavity up to a maximum detuning of $3.2\,\mathrm{rad}$, at which point the experimentally accessible spectral bandwidth reaches $1.3\,\mathrm{THz}$. This is significantly narrower than in simulation, owing to amplified spontaneous emission (ASE) obscuring the wings of the spectrum. In addition, the intracavity gain profile peaks near $1550\,\mathrm{nm}$, while the pump operates at $1554\,\mathrm{nm}$. As a result, blue-detuned components (closer to the gain maximum) experience higher amplification than red-detuned components. This behavior makes the Kelly sidebands closer to $1550\,\mathrm{nm}$ appear more pronounced than those farther from $1550\,\mathrm{nm}$.

In Fig.~\ref{fig:snrvsM}, a 10-node reservoir spanning the spectral range $[1550,1560]\,\mathrm{nm}$ is configured using bandpass filters with a bandwidth of $120\,\mathrm{GHz}$. We characterized the Signal-to-Noise Ratio (SNR) per node, with the central $70\,\mathrm{GHz}$ around the pump wavelength excluded. The SNR of the induced power fluctuations is defined as the ratio between the calculated signal power ($P_{\mathrm{signal}}$) and the measured noise power ($P_{\mathrm{noise}}$). The power measured within a given time window is computed as the mean-square of the recorded voltage after DC-offset removal: $\Delta V(t)\,=\,V(t)-\langle V(t)\rangle$, so that $P\,=\,\langle \Delta V^{2}(t)\rangle$. Our methodology assumes that the noise is additive and uncorrelated with the signal, such that $P_{\mathrm{total}}\,=\,P_{\mathrm{signal}} + P_{\mathrm{noise}}$. The noise power $P_{\mathrm{noise}}$ is evaluated from an initial segment of the time trace where the cavity is unperturbed. The signal power is then obtained by subtracting the noise power from the total power measured during the perturbed state. The SNR is finally calculated as \(\mathrm{SNR}_{\mathrm{dB}}\,=\,10\log_{10}\!\left(\frac{P_{\mathrm{signal}}}{P_{\mathrm{noise}}}\right).\) The SNR increases with the modulation depth $M$, and is smallest (i) near the pump wavelength, where the relative spectral variations are minimal and ASE noise dominates, and (ii) in the spectral tails, where the optical power is low.

\section{Experimental Results}
\label{sec:EXP_results}

In this section, we experimentally benchmark the cavity soliton as a reservoir computer. The choice of experimental parameters is dictated by the following considerations. First, recall (see Fig.~\ref{fig:M_Q}) that for $P_{\mathrm{in}}\,=\,200\,\mathrm{mW}$ (a typical experimental value), numerical simulations predict an optimal repetition factor $q\,=\,10$ and a very weak modulation amplitude. However, for such weak modulation amplitudes, the signal is not experimentally detectable. We therefore increase $M$ to obtain a measurable soliton response, and use a correspondingly larger $q$ in order to maintain stability of the soliton. In addition, Raman scattering not only leads to soliton self-frequency shift, but it also reduces the maximum achievable detuning compared to the theoretical prediction of the models described in Sec.~\ref{sec:models} (which neglects Raman effects~\cite{Yi2016}). At high detunings ($\delta \ge 2\,\mathrm{rad}$), the system approaches its stability limit. Perturbing the soliton at high detunings can easily push it outside the stable operating region. While at low detunings ($\delta \le 0.5\,\mathrm{rad}$), the cavity is difficult to stabilize in practice, leading to the loss of CSs during the experiment. Therefore, the detuning values reported in the following sections represent optimized values that maintain soliton stability under strong phase perturbation. We tested the experimental system on the XOR task and the chaotic Hénon map.

\subsection{XOR benchmark}
\label{subsec:XOR}
\begin{figure}[htbp]
	\centering
	\begin{subfigure}[t]{0.52\textwidth}
		\centering
		\vspace{0pt}
		\includegraphics[width=\textwidth]{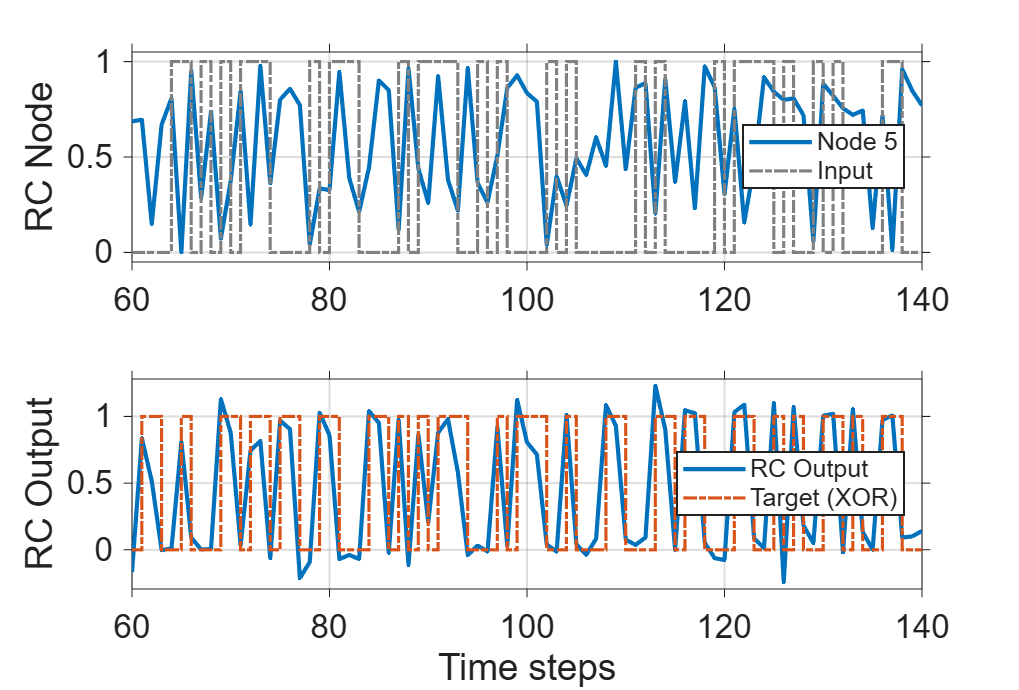}
		\caption{Reservoir node/output versus input/target signal}
		\label{fig:RC_XOR}
	\end{subfigure}
	\hspace{.5cm}
	\begin{subfigure}[t]{0.4\textwidth}
		\centering
		\vspace{0pt}
		\includegraphics[width=\textwidth]{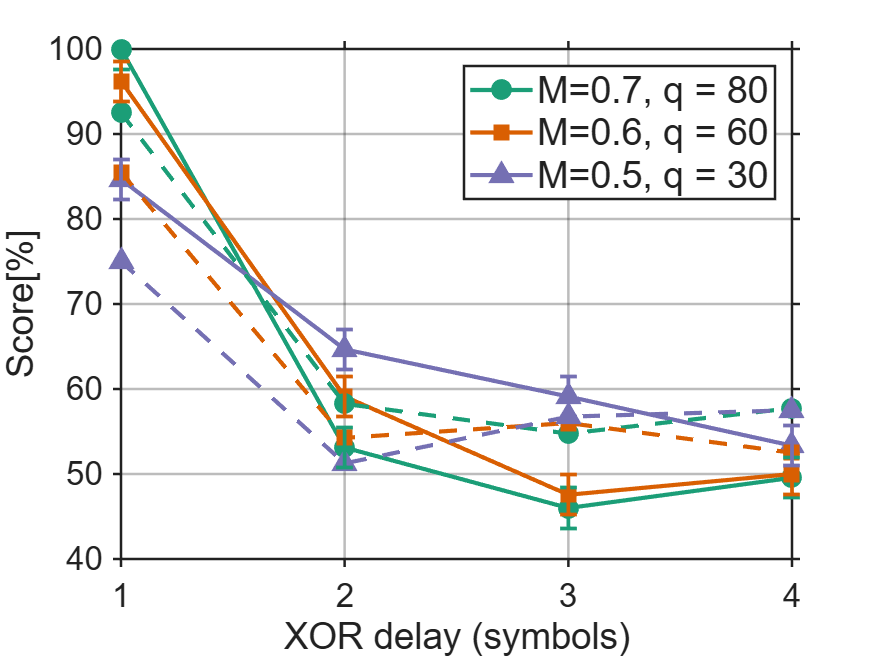}
		\caption{XOR accuracy as a function of delay between symbols}
		\label{fig:XOR_results}
	\end{subfigure}
	\caption{ Experimental results on XOR task.
		(a) The upper panel shows the response of the fifth reservoir node (blue trace) to the binary input sequence (black trace). The lower panel compares the final reservoir output (blue trace) with the target XOR sequence (orange trace). The data corresponds to a modulation strength of $M\,=\,0.7$ and repetition factor of $q\,=\,80$.
		(b) Accuracy as a function of input delay for different values of \(M\). Solid lines correspond to experimental results, while dashed lines show numerical simulations.}
	\label{fig:XOR}
\end{figure}

To implement the XOR benchmark, the detuning was fixed at $\delta\,=\,1.5\,\mathrm{rad}$ and the input power set to $P_{\mathrm{in}}\,=\,200\,\mathrm{mW}$. At the readout stage, $N\,=\,10$ frequency channels each with bandwidth of $120\,\mathrm{GHz}$ are applied. Three modulation strengths were used $M\,=\,\{0.5,0.6,0.7\}$ with repetition factor $q\,=\,30$, $60$, and $80$ respectively. (For larger modulation strength, the repetition factor needed to be increased; otherwise the solitons were destroyed).

The experimental results are presented in Fig.~\ref{fig:XOR}. A sequence of $K\,=\,1000$ random binary symbols was injected into the reservoir. The first 100 symbols were used as a washout period, the remaining ones were split into two equal parts: one for training and the other for testing. The system was trained to compute the XOR between the current and delayed inputs. Classification score was computed as the ratio of correctly predicted outputs to the total number of test symbols. 

Figure~\ref{fig:RC_XOR} illustrates how the system works for a configuration with a modulation strength of $M\,=\,0.7$ and a repetition factor of $q\,=\,80$. The solid lines in Fig.~\ref{fig:XOR_results} show the system performance as a function of delay between inputs. The higher values of $M\,=\,0.6$ and $M\,=\,0.7$ perform better for a delay of $1$, while the lower value of $M\,=\,0.5$ performs worse for a delay of $1$, but better for longer delays between inputs. The numerical results are shown as dashed lines in Fig.~\ref{fig:XOR_results}. In these numerical simulations, the spectrum is truncated so that each node has a bandwidth of $120\,\mathrm{GHz}$, and independent random noise is added at the reservoir output such that the node-wise SNR is $25\,\mathrm{dB}$.

\subsection{Hénon map}
\label{subsec:Henon}
\begin{figure}[htbp]
	\centering
	\begin{subfigure}[b]{0.32\textwidth}
		\vspace{0pt}
		\includegraphics[width=\linewidth]{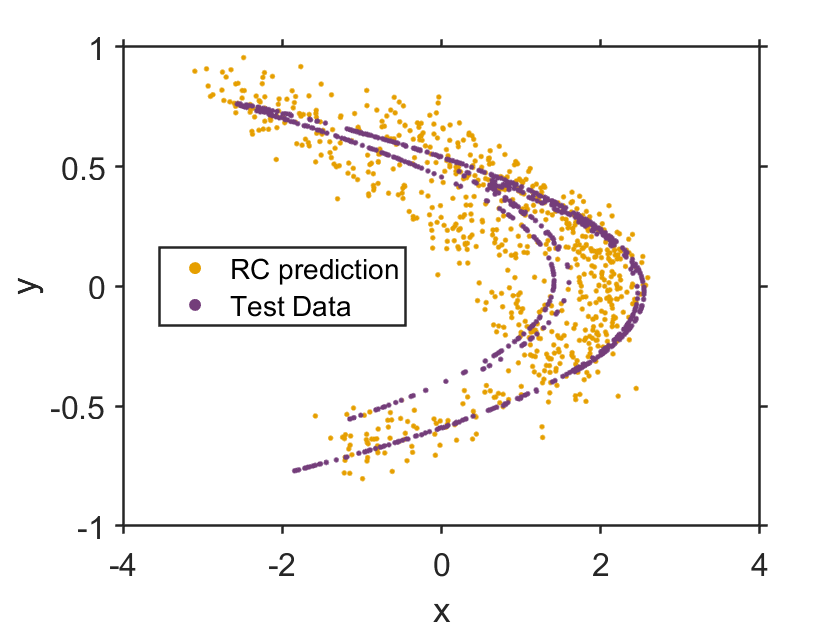}
	\end{subfigure}
	\hfill
	\begin{subfigure}[b]{0.66\textwidth}
		\vspace{0pt}
		\includegraphics[width=\linewidth]{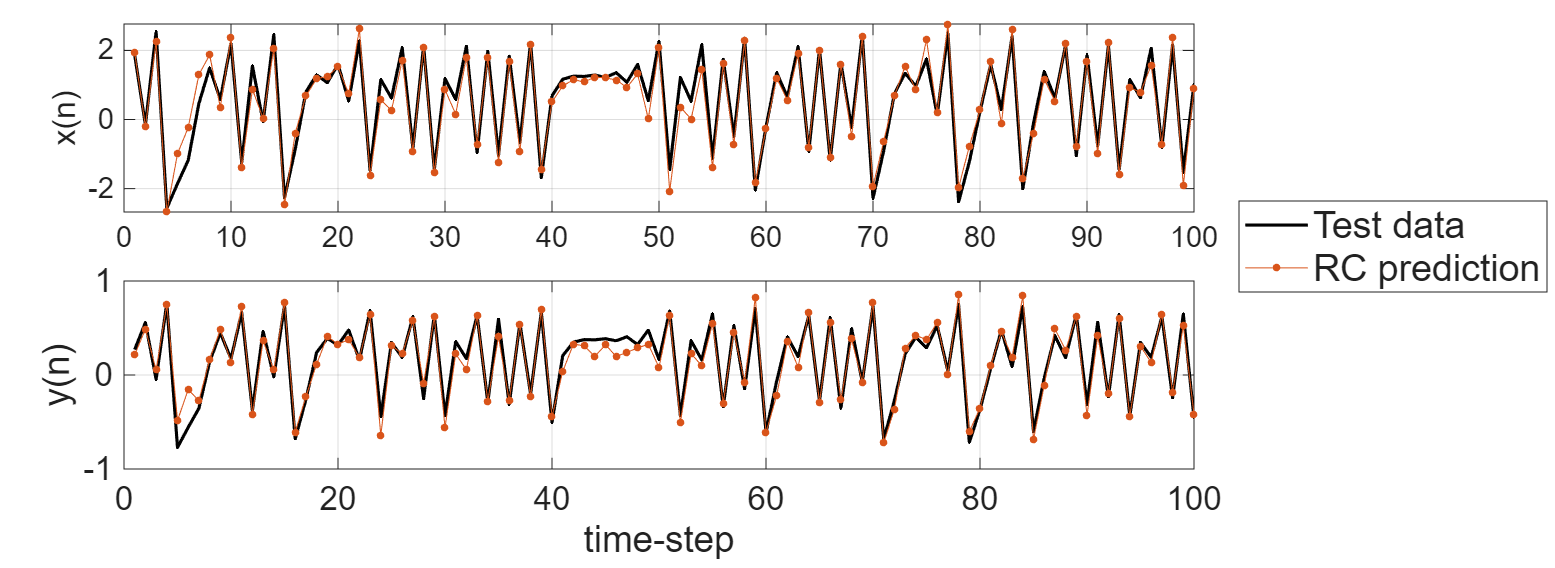}
	\end{subfigure}
	\caption{One-step-ahead prediction of the Hénon map using the cavity-soliton reservoir computer. Left: Two-dimensional Hénon map attractor (violet) and reservoir prediction (yellow). Right: 100-step segment of the test data (black line) and reservoir prediction (orange markers) are shown for the $x$ (upper panel) and $y$ (lower panel) coordinates.}
	\label{fig:henonmap}
\end{figure}
The Hénon map is a two-dimensional chaotic system defined by the coupled nonlinear equations 
\(x_{t+1} = 1 - a x_t^2 + y_t\) and \(
y_{t+1} = b x_t\), 
with parameters $a\,=\,1.4$ and $b\,=\,0.3$~\cite{Henon1976}. Its quadratic nonlinearity, intrinsic one-step memory, and sensitivity to initial conditions and noise make it a good test for analogue neuromorphic computing. The goal of this benchmark is to test the reservoir’s ability to predict the next state of the Hénon system from its current state. Starting from the initial conditions $x_0\,=\,0$ and $y_0\,=\,0$, we generated a sequence of 8000 $(x,y)$ symbols from the Hénon map and fed the $x$ coordinates into a reservoir. This benchmark is significantly more challenging than XOR bit classification, requiring stronger perturbations to the soliton state. To achieve this, we increase the cavity detuning to $\delta\,=\,2\,\mathrm{rad}$ and the input power to $P_{\mathrm{in}}\,=\,250\,\mathrm{mW}$, pushing the system closer to its stability limit. Additionally, we increase the number of spectral filters to $N\,=\,20$ output nodes, each with a bandwidth of $60\,\mathrm{GHz}$, to capture the finer spectral variations induced by the perturbations. The system is configured with modulation strength $M\,=\,0.7$, and repetition factor $q\,=\,35$. Each node measurement was repeated ten times and the results were averaged to reduce the impact of noise. Subsequently, we applied a built-in MATLAB Savitzky--Golay filter (quadratic, window\,=\,55 samples) to suppress high-frequency noise while preserving the temporal profile of the reservoir states. The reservoir's readout was then trained on 80\% of the data to predict the next step in the time-series. Figure~\ref{fig:henonmap} shows the resulting prediction, which achieved a normalized root-mean-square error (NRMSE) of $0.3$, while a linear reservoir would achieve a NRMSE of $0.44$. Using the same experimental parameters (except with $q\,=\,70$ and no added noise), we obtained an NRMSE of 0.28 numerically.

These experimental results show that the cavity-soliton reservoir computer can carry out information processing tasks that require both memory and nonlinearity.

\section{Conclusion and outlook}
\label{sec:outlook}	
Harnessing the frequency degree of freedom could provide a scalable route for high-speed photonic information processing. This was first demonstrated for reservoir computing using electro-optic combs and a linear cavity in Ref.~\cite{Butschek2022}. Recent studies notably, Refs.~\cite{Shishavan2025, Cuevas2025} and the present contribution aim to improve on this by leveraging the Kerr nonlinearity and the resulting nonlinear cavity response.

The numerical study in Ref.~\cite{Shishavan2025} employs a frequency multiplexing approach within the modulational-instability (MI) driven chaotic-comb regime. This raises the question, which deserves further investigation, of whether operating in a chaotic regime compromises the reproducibility of the reservoir output.

Both Ref.~\cite{Cuevas2025} and the present work use cavity solitons for reservoir computing with input information encoded in the phase of the driving field. However they differ in several key aspects: First, regarding the optical platform, while Ref.~\cite{Cuevas2025} utilizes a micro-ring resonator producing a frequency comb with 100~GHz spacing, our fiber-cavity implementation generates cavity solitons with a much denser, near-continuous, spectrum with a line spacing of 3.3~MHz. This platform distinction directly impacts the physical dynamics; In our fiber cavity, various lumped perturbations—such as the input coupler—induce periodic soliton reshaping and the emission of radiative waves. This necessitates the use of the Ikeda map for realistic modeling and leads to the excitation of Kelly sidebands. In contrast, the near-uniformity of micro-ring resonators drastically reduces soliton reshaping, causing Kelly sidebands to be strongly suppressed and resulting in almost identical behavior between the Ikeda map and the mean-field LLE model. Our numerical simulations show that, in our parameter regime, the Ikeda map predicts better performance on reservoir computing tasks than the LLE. We attribute this performance boost to the spectral interference between Kelly sidebands and cavity solitons which enhances the temporal diversity across the reservoir nodes. Second, the two systems also operate in substantially different parameter regimes. When describing soliton dynamics, the LLE can be rescaled into a dimensionless form that depends on only two parameters: the normalized detuning $\Delta = \frac{2\delta}{\Lambda}$ and the normalized drive $X = \frac{8\gamma P_{\text{in}}\theta_{\text{in}}L}{\Lambda^3}$. The numerical simulations in Ref.~\cite{Cuevas2025} fix $X=32$ and perform benchmarks near the lower edge of the soliton branch ($\Delta \simeq 12$) with modest excursions to 16 and 18. In contrast, our implementation (see for instance Sec.~\ref{subsec:XOR}) utilizes a much higher normalized power $X \simeq 385$ and a stabilized detuning of $\Delta = 100$, with excursions reaching $\Delta = 146$. This increased modulation depth drives the reservoir further into the nonlinear regime. Lastly, in terms of post-processing, while we focus on the raw performance of the fiber-based reservoir, Ref.~\cite{Cuevas2025} employs a technique involving random inter-mode delays on the comb intensity traces to enhance performance.

We expect that soliton-based reservoir computers can be further improved by leveraging several already demonstrated techniques. First, as shown in~\cite{Butschek2022}, frequency-based reservoir computers can be naturally extended to include an analog output layer, a feature that is generally more challenging to implement in other approaches. Second, the soliton dynamics can be further enriched, for instance by enhancing the Kelly sidebands in fiber cavities, as pointed out in our numerical simulations. Third, the effective dimensionality of the reservoir can be expanded by dividing the signal acquired during the $q$-roundtrips into several temporal sub-windows, as proposed in ~\cite{Zhong2021}. Finally, by improving the conversion efficiency from the continuous-wave (CW) pump to frequency-comb lines~\cite{Jang2021,Li2022}, together with achieving flat-top comb profiles~\cite{Xue2019,Helgason2023,Boggio2022,Zhang2024,Kondratiev2023}, should improve energy efficiency and increase the number of usable reservoir nodes.

\paragraph{Funding}
We acknowledge funding by the Fonds National de la Recherche Scientifique – FNRS through projects EQP n° 40021523, EOS O.0019.22 (Photonic Ising Machines) and EOS n°40007560 (PULSE), and the ERC project HIGHRES (grant agreement No 101125625)
\paragraph{Acknowledgment}
Authors would like to thank Chee Wei Wong, Hangbo Yang, Tristan Melton, François Leo, Georges Semaan, Haftamu Berhe, and Jesús Yelo-Sarrión for stimulating discussions and technical help.
\paragraph{Disclosures}
The authors declare no conflicts of interest.
\paragraph{Data availability}
Data underlying the results presented in this paper are not publicly available at this time but may be obtained from the authors upon reasonable request.
\paragraph{Supplementary material}
See Supplementary material for a detailed derivation of cavity-soliton dynamics under slow-time phase modulation of the driving field.

\appendix
\section{Derivation of the Reduced Model}
In this supplemental appendix, we derive a reduced set of equations of motion for the cavity-soliton amplitude and phase Eqs.~\eqref{eq:EOMs}. Starting from the Lugiato--Lefever equation Eq.~\eqref{eq:LLE}, using a Lagrangian formulation, we obtain a reduced Lagrangian and the corresponding Euler--Lagrange equations. In this reduced description, cavity solitons exhibit no frequency shift and no temporal drift.

Considering that the phase of the driving field undergoes \textit{slow-time} changes $\varphi(t)$, as described by Eq.~\eqref{eq:LLE} of the primary manuscript, we make the substitutions
$E^{\prime}_{\mathrm{in}}=E_{\mathrm{in}} e^{i \varphi(t)}$ and
$E^{\prime}=E e^{i \varphi(t)}$.
The Lugiato--Lefever equation then becomes

\begin{equation}
	t_R\left(\frac{\partial E^{\prime}(t,\tau)}{\partial t} +
	iE^{\prime}(t,\tau)\frac{d\varphi(t)}{dt}\right) =
	\left(-\frac{\Lambda}{2} - i\delta - i\frac{\beta_2 L}{2}\partial^2_{\tau} + i\gamma L |E^{\prime}(t,\tau)|^2\right)E^{\prime}
	+ \sqrt{\theta_{\mathrm{in}}}\, E^{\prime}_{\mathrm{in}}.
	\label{eq:app1}
\end{equation}

A Lagrangian density whose Euler--Lagrange equation reproduces Eq.~\eqref{eq:app1} is given by
\begin{multline}
	\mathcal{L}
	=e^{\Lambda t/t_R}\Big(
	\frac{t_R}{2}\left(E^{\prime}\partial_tE^{\prime*}-E^{\prime*}\partial_tE^{\prime}\right)
	- i\delta|E^{\prime}|^2
	+ i\frac{\beta_2L}{2}|\partial_{\tau}E^{\prime}|^2
	+ i\frac{\gamma L}{2}|E^{\prime}|^4  \\
	+ \sqrt{\theta_{\mathrm{in}}}\left(E^{\prime*}E^{\prime}_{\mathrm{in}} - E^{\prime}E^{\prime*}_{\mathrm{in}}\right)
	- it_R\partial_t \varphi \,|E^{\prime}|^2
	\Big),
	\label{eq:LagDens}
\end{multline}
where loss is incorporated in the Lagrangian by multiplying it with a time-dependent exponential factor. The general soliton \textit{ansatz} takes the form
\begin{equation}
	E^{\prime}_{CS}(t,\tau) = \eta(t)\,\mathrm{sech}\!\left(\eta(t)\,\frac{\tau-\tau_0(t)}{\sqrt{-\beta_2/\gamma}}\right)
	\exp\!\left\{\, i\big[\phi(t) - \kappa(\tau - \tau_0(t))\big] \right\},
	\label{eq:Solansatz}
\end{equation}
where $\eta(t)$ is the soliton amplitude, $\tau_0(t)$ its temporal position, $\kappa$ the frequency shift, and $\phi(t)$ the soliton phase. Inserting Eq.~\eqref{eq:Solansatz} into Eq.~\eqref{eq:LagDens} and integrating over the fast-time coordinate $\tau$ yields the reduced Lagrangian:
\begin{multline}
	L = e^{\Lambda t/t_R}\Bigg[
	2\eta(t)\left(\delta + t_R\left(\dot{\phi}(t) + \kappa \dot{\tau_0}(t) + \dot{\varphi}(t)\right)\right)
	+\eta(t)\gamma\,L\kappa^2 - \frac{L\gamma\eta^3(t)}{3}  \\
	+ 2\sqrt{\theta_{\mathrm{in}} P_{\mathrm{in}}}\,\pi\,\mathrm{sech}\left(\frac{\kappa\pi}{2\eta}\right)\sin(\phi(t))
	\Bigg].
	\label{eq:app4}
\end{multline}

The Euler--Lagrange equations for the dynamical variables $\kappa$ and $\tau_0$ yield
\begin{subequations}
	\begin{align}
		&\kappa\frac{d}{dt}\left(e^{\Lambda t/t_R}t_R\eta(t)\right)=0,
		\label{eq:EOM_tau0}\\
		&t_R\dot{\tau}_0(t)=\frac{\pi ^2 \sqrt{\theta_{\mathrm{in}}  P_{\mathrm{in}}} \sin (\phi (t)) \tanh \left(\frac{\pi  \kappa (t)}{2 \eta (t)}\right) \mathrm{sech}\left(\frac{\pi  \kappa (t)}{2 \eta (t)}\right)}{2\eta^2 (t)} - \gamma L\kappa .
		\label{eq:EOM_kappa}
	\end{align}
\end{subequations}

Since a cavity soliton does not exponentially decay in time, Eq.~\eqref{eq:EOM_tau0} implies that $\kappa=0$. Substituting $\kappa=0$ into Eq.~\eqref{eq:EOM_kappa} implies that $\tau_0$ is constant. The Euler--Lagrange equations for the amplitude and phase parameters are then found to be Eqs.~\eqref{eq:EOMs} of the primary manuscript. In the absence of perturbations on the driving beam, these equations of motion admit approximate soliton solutions previously mentioned in Refs.~\cite{coen2013,Wabnitz1993}.

\end{document}